\documentclass[12pt]{article}
\usepackage{amsmath}
\usepackage{graphicx}
\usepackage{enumerate}
\usepackage{natbib}
\usepackage{url} 
\usepackage{float}
\usepackage{amsmath,amssymb,mathtools,amsthm,mathrsfs,bbm}
\usepackage{algorithm}
\usepackage{algorithmic}
\usepackage{tikz}
\usepackage{graphicx}
\usepackage{epstopdf}
\usepackage{subcaption}
\usepackage{pgfplots}
\pgfplotsset{compat=1.16}
\usepgfplotslibrary{colormaps}
\RequirePackage[colorlinks,citecolor=blue,urlcolor=blue]{hyperref}
\usepackage{epstopdf,color}
\usepackage{booktabs, float} 
\usepackage{multirow}

\newtheorem{assumption}{Assumption}
\newtheorem{theorem}{Theorem}
\newtheorem{remark}{Remark}

\newcommand{\bSigma}{\boldsymbol{\Sigma}}

\newcommand{\E}{\mathbb{E}}

\newcommand{\bB}{\mathbf{B}}

\newcommand{\bI}{\mathbf{I}}

\newcommand{\bM}{\mathbf{M}}

\newcommand{\bR}{\mathbf{R}}
\newcommand{\bS}{\mathbf{S}}

\newcommand{\bU}{\mathbf{U}}
\newcommand{\bV}{\mathbf{V}}
\newcommand{\bW}{\mathbf{W}}
\newcommand{\bX}{\mathbf{X}}

\newcommand{\be}{\mathbf{e}}

\newcommand{\bu}{\mathbf{u}}

\newcommand{\bv}{\mathbf{v}}
\newcommand{\bw}{\mathbf{w}}
\newcommand{\bx}{\mathbf{x}}
\newcommand{\by}{\mathbf{y}}

\newcommand{\sign}{\mathrm{sign}}
\newcommand{\diag}{\mathrm{diag}}

\renewcommand{\bar}{\overline}
\usepackage{dsfont}

\definecolor{Red1}{RGB}{142, 0, 40}
\definecolor{Red2}{RGB}{189, 60, 51}
\definecolor{Orange1}{RGB}{228, 148, 90}
\definecolor{Orange2}{RGB}{249, 198, 118}
\definecolor{Yellow}{RGB}{251, 236, 171}
\definecolor{Blue1}{RGB}{255, 255, 255}
\definecolor{Blue2}{RGB}{209, 226, 239}
\definecolor{Blue3}{RGB}{127, 169, 205}
\definecolor{Blue4}{RGB}{82, 124, 180}
\definecolor{Blue5}{RGB}{63, 96, 163}
\definecolor{Blue6}{RGB}{48, 50, 125}

\newcommand{\blind}{1}

\floatname{algorithm}{Algorithm} 

\begin{document}

\def\spacingset#1{\renewcommand{\baselinestretch}%
{#1}\small\normalsize} \spacingset{1}


\if1\blind
{
  \title{\bf Debiased distributed PCA under high dimensional spiked model}
  \author{Weiming Li\\
    \small School of Statistics and Management, Shanghai University of Finance and Economics \\
     Zeng Li
     \thanks{Corresponding author, email: liz9@sustech.edu.cn}
  \hspace{.2cm}\\
    \small Department of Statistics and Data Science, Southern University of Science and Technology\\
    Siyu Wang \\
    \small School of Mathematical Sciences, Beijing Normal
University\\
    Yanqing Yin \\
    \small School of Statistics and Data Science, Nanjing Audit University\\
    Junpeng Zhu\\
    \small Department of Statistics and Data Science, Southern University of Science and Technology\\
    }
  \maketitle
\renewcommand{\thefootnote}{} 
\footnote{All authors equally contributed to this work and are listed in the alphabetic order.}
} \fi

\if0\blind
{
  \bigskip
  \bigskip
  \bigskip
  \begin{center}
    {\LARGE\bf Debiased distributed PCA under high dimensional spiked model}
\end{center}
  \medskip
} \fi

\bigskip
\begin{abstract}
We study  distributed principal component analysis (PCA) in high-dimensional settings under the spiked model. In such regimes, sample eigenvectors can deviate significantly from population ones, introducing a persistent bias. Existing distributed PCA methods are sensitive to this bias, particularly when the number of machines is small. Their consistency typically relies on the number of machines tending to infinity. We propose a debiased distributed PCA algorithm that corrects the local bias before aggregation and incorporates a sparsity-detection step to adaptively handle sparse and non-sparse eigenvectors. Theoretically, we establish the consistency of our estimator under much weaker conditions compared to existing literature. In particular, our approach does not require symmetric innovations and only assumes a finite sixth moment. Furthermore, our method  generally achieves smaller estimation error, especially when the number of machines is small. Empirically, extensive simulations and real data experiments demonstrate that our method consistently outperforms existing distributed PCA approaches. The advantage is especially prominent when the leading eigenvectors are sparse or the number of machines is limited. Our method and theoretical analysis are also applicable to the sample correlation matrix.
\end{abstract}

\noindent%
\vfill

\newpage
\spacingset{1.9} 

\section{Introduction}
Distributed principal component analysis (PCA) is crucial for handling massive, distributed datasets. Since data is often scattered across different locations and cannot be easily centralized for storage and computation, traditional PCA methods face significant challenges in high-dimensional, large-scale settings. Distributed PCA enables efficient computation by extracting principal components locally and aggregating information across subsets, reducing communication costs while improving computational efficiency. This makes it particularly valuable in scenarios with privacy constraints or limited computational resources. Therefore, developing efficient and robust distributed PCA methods is essential for large-scale data analysis.

Distributed PCA typically falls into two categories, depending on how the data is partitioned: across features (vertical) or across samples (horizontal). In the ``vertical" regime, the features are divided into several parts, with each storage node having full access to the sample but only a subset of the features. 
This regime has gained attention
 in signal processing and sensor networks. For example, \cite{kargupta2001distributed}
 proposed the collective PCA algorithm, which performs local PCA and communicates reduced data and eigenvectors for global analysis.
\cite{bertrand2014distributed}
  introduced a distributed adaptive approach in sensor networks, where each node iteratively exchanges local estimates with its neighbors to collaboratively approximate the global principal components.
\cite{schizas2015distributed} reformulated PCA as a separable constrained minimization problem, enabling efficient eigenspace estimation and denoising in sensor networks.
More recent approaches have focused on federated learning settings, see 
\cite{cheung2021vertical,shen2023fadi} etc.

In contrast, the ``horizontal" approach retains all features in each subserver, while the samples are distributed across different machines. 
Specifically, consider $N$ random samples $\{\bx_i\}_{i=1}^N\subseteq \mathbb{R}^p$ scattered across $m$ machines, where  $N=\sum_{\ell=1}^m n_\ell$ and the $\ell$-th machine  storing $n_\ell$ samples. Let $\mathbf{\Sigma}^{(\ell)}=\sum_{i=1}^p \lambda_i^{(\ell)} \mathbf{u}_i^{(\ell)} \mathbf{u}_i^{(\ell)\top} $ denote the population covariance matrix on machine $\ell$. For a given $K$, assume that $\left\{\mathbf{\Sigma}^{(\ell)}\right\}_{\ell=1}^m$ share the same projection matrix of the top $K$ eigenspace, i.e., $\sum_{i=1}^K \mathbf{u}_i^{(\ell)} \mathbf{u}_i^{(\ell)\top}=\sum_{i=1}^K \mathbf{u}_i\mathbf{u}_i^{\top}$ for all $1\leq \ell\leq m$. The 
objective is to estimate the common leading eigenspace $\sum_{i=1}^{K}\mathbf{u}_i \mathbf{u}_i^{\top}$ under this distributed data setting. We focus on this setting in this paper.


The horizontal  regime  has attracted significant research attention. A typical approach is to perform eigen-decomposition locally on each machine and then aggregate the results. 
\citet{qu2002principal} considered a setting where machines may have different population means and proposed a decomposition of the global covariance matrix into within-location and between-location components. Their distributed PCA approach retains the global between-location structure while approximating the within-location component using the average of local principal components, thereby enabling robust eigenspace estimation under heterogeneous mean settings.
\cite{fan} proposed to compute the top $K$ sample eigenvectors $\{\hat{\mathbf{u}}_i^{(\ell)}\}_{i=1}^{K}$ of sample covariance matrix locally on each machine and then  perform PCA based on the aggregated data $(1 / m) \sum_{\ell=1}^m \sum_{i=1}^K \hat{\mathbf{u}}_i^{(\ell)} \hat{\mathbf{u}}_i^{(\ell) \top}$.
To handle heavy-tailed data, \cite{he2024distributed} extended the framework of \cite{fan} to  sample multivariate Kendall’s tau matrix. 
\cite{li2025two} proposed a two-round distributed PCA algorithm that introduces an additional refinement step, where each local eigenspace is projected onto the aggregated subspace $(1 / m) \sum_{\ell=1}^m \sum_{i=1}^K \hat{\mathbf{u}}_i^{(\ell)} \hat{\mathbf{u}}_i^{(\ell)\top}$ to improve global alignment.
Moreover, \cite{li2024knowledge} adopted a transfer learning perspective that allows variation in the top $K$ leading eigenspace. They decompose each local eigenspace into a shared component and a private component, enabling a more flexible and principled modeling of heterogeneity.

Existing literature mostly focuses on algorithm design, while offering little discussion on the relationship between the dimension $p$ and the local sample size $n_{\ell}$ on each machine. However, this relationship 
plays a crucial role in the performance of PCA. As shown in \cite{MR145620}, when $p$
is fixed and $n_{\ell} \rightarrow \infty$,  the local sample eigenvectors are consistent estimators of the true eigenvectors. In this regime, even a simple averaging of local PCA results can yield consistent estimators. Any number of machines 
$m$ can achieve reliable estimation. However, under the high-dimensional regime where 
$$
n_{\ell} \rightarrow \infty, ~p\rightarrow \infty, ~p / n_{\ell} \rightarrow c_{\ell} \in(0, \infty),\quad 
1\leq \ell\leq m,
$$
the local sample eigenvectors $\hat\bu_{i}^{(\ell)}$ become biased and deviate from the population eigenvectors $\bu_i$ by a non-zero angle. The asymptotic limit of the inner product between 
$\hat{\mathbf{u}}_i^{(\ell)}$ and  $\bu_i$ under the spiked model
has been  extensively studied in 
\cite{Paul2007,MR2589886,MR2782201,couillet2022random,yinpca}. Regrettably, the deviation between $\hat{\mathbf{u}}_i^{(\ell)}$ and $\bu_i$ has been largely overlooked in existing distributed PCA methods.
When the number of machines $m$ is small, existing methods exhibit noticeable numerical bias.
They rely on increasing the number of machines $m \to \infty$ and the assumption of symmetric innovation of data distribution to eliminate estimation bias and achieve global consistency, see \cite{fan,he2024distributed,li2025two}.

In this paper, we propose a bias-corrected method based on the spiked model to improve the performance of the global estimator. Specifically, we correct the bias of the sample eigenvectors $\hat{\mathbf{u}}_i^{(\ell)}$ locally
on each machine 
and then aggregate the debiased eigenspace. 
We relax the sub-Gaussian and symmetric innovation assumptions in prevailing literature and establish consistency under the weaker condition of a finite sixth moment of data distribution as $m\to \infty$.
When  $m$ is small, our estimator $\sum_{i=1}^K \check{\mathbf{u}}_i \check{\mathbf{u}}_i^{\top}$ of $\sum_{i=1}^K \mathbf{u}_i \mathbf{u}_i^{\top}$ demonstrates significantly superior numerical performance,  achieving a much smaller statistical error $\rho=\left\|\sum_{i=1}^K \check{\mathbf{u}}_i \check{\mathbf{u}}_i^{\top}-\sum_{i=1}^K \mathbf{u}_i \mathbf{u}_i^{\top}\right\|_F$. Moreover, our algorithm includes an additional step for detecting and handling sparsity in the eigenvectors.
When the population eigenvectors are sparse, our proposed estimator achieves consistency (i.e., $\rho \to 0$) for any fixed number of machines $m$, and attains substantially smaller $\rho$ values compared to existing methods. Lastly, our theoretical guarantees and empirical results are valid for both sample covariance and correlation matrices. 
To further illustrate the practical effectiveness of our method, we conducted experiments on the  MiniBooNE dataset 
which is collected from tabular data benchmark. 
We apply the distributed PCA algorithms to the training set to estimate the top-$K$ eigenspace, and then evaluate the Average Information Preservation Ratio (AR) on the test set to quantify how much of the original information is retained by the estimated principal components.
Experimental results show that our method generally achieves higher AR values, indicating its superior effectiveness and stronger capability in preserving the informative structure of the data.

The remainder of the paper is organized as follows. Section \ref{sec:2} introduces some preliminary knowledge on the spiked model. In Section \ref{sec:3}, we describe the proposed distributed algorithm. Section \ref{sec:4} analyzes its asymptotic properties. Section \ref{sec:5}  presents simulation studies. Real data analysis is shown in Section \ref{sec:6}.
Finally, all technical proofs are detailed in the Supplementary Material.

\section{Preliminaries on the 
 spiked model}\label{sec:2}
\subsection{Notation}
We first collect the notation that will be used in this paper. We have $N$ random
samples $\{\bx_i\}_{i=1}^N\subseteq \mathbb{R}^p$ scattered across $m$ machines, where $N=\sum_{\ell=1}^m n_\ell$ and the $\ell$-th machine storing $n_\ell$ samples. 
For ease of exposition, we consider the case where the $N$ samples share a common population covariance matrix $\bSigma$. Our method and theoretical analysis can be extended to a more general setting where the covariance matrices $\bSigma^{(\ell)}$ differ across machines but share the same top-$K$ eigenspace.
 As for common $\bSigma$, we have $\boldsymbol{\Sigma}=\sum_{i=1}^p \lambda_i \mathbf{u}_i \mathbf{u}_i^{\top}$ by spectral decomposition, where 
$\lambda_1 \geq \lambda_2 \geq \ldots \geq \lambda_p$. For the $\ell$-th machine ($1 \leq \ell \leq m$), let $\mathbf{X}_{n{\ell}}^{(\ell)} \in \mathbb{R}^{p \times n_{\ell}}$ denote the local data matrix, and $\mathbf{S}_{n{\ell}}^{(\ell)}$ be the corresponding sample covariance matrix. 
$\left\{\hat{\lambda}_i^{(\ell)}\right\}_{i =1}^p$ and $\left\{\hat{\mathbf{u}}_i^{(\ell)}\right\}_{i=1}^p$
are the corresponding sample eigenvalues and eigenvectors of $\mathbf{S}_{n_{\ell}}^{(\ell)}$. 
Our goal is to estimate the top $K$ eigenspace $\sum_{i=1}^K \mathbf{u}_i \mathbf{u}_i^{\top}$ through $\mathbf{S}_{n_{\ell}}^{(\ell)}$.

\subsection{Spiked model}
In this paper, we assume that the population covariance matrix $\bSigma=\sum_{i=1}^p \lambda_i \mathbf{u}_i \mathbf{u}_i^{\top}$ follows the well-established spiked  model in random matrix theory, see \cite{Johnstone01,2008Central,MR4474488}, etc. Specifically, we assume the top $K$ eigenvalues $\lambda_1>\lambda_2>\ldots>\lambda_K$ (spiked eigenvalues) are well-separated from the remaining eigenvalues $\lambda_{K+1}\geq\lambda_{K+2}\geq\ldots\geq\lambda_p$ (bulk eigenvalues).
A famous example of the spiked model is the factor model.

\begin{figure}[H]
     \centering
    \begin{subfigure}{0.45\textwidth}
        \centering
\includegraphics[width=\textwidth]{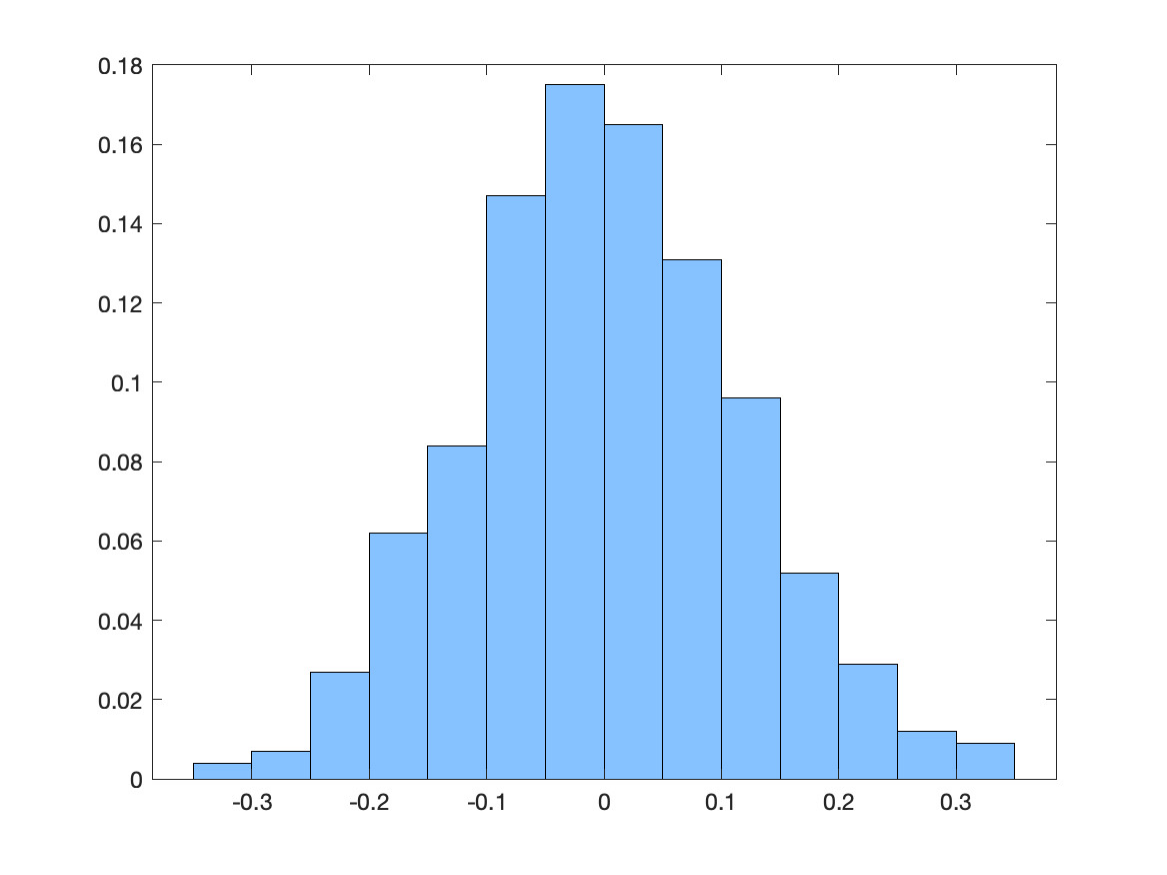} 
        \caption{Histogram of $\hat{\mathbf{u}}_1^{\top} \mathbf{u}_1$(non-spiked model)}
\end{subfigure}
     \begin{subfigure}{0.4\textwidth}
        \centering
   \includegraphics[width=\textwidth]{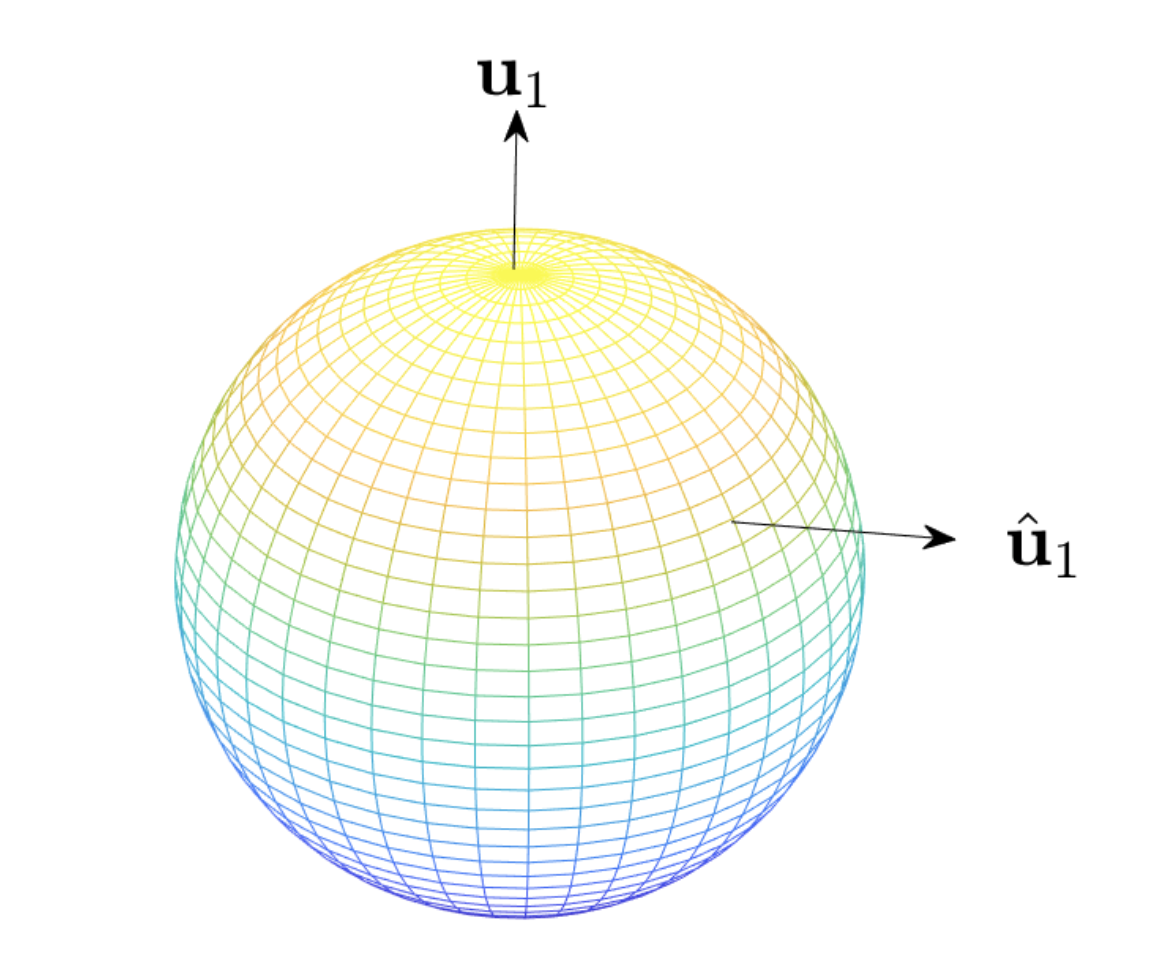} 
        \caption{$\hat\bu_1$ under the non-spiked model}
    \end{subfigure}\\
        \centering
    \begin{subfigure}{0.5\textwidth}
        \centering
\includegraphics[width=\textwidth]{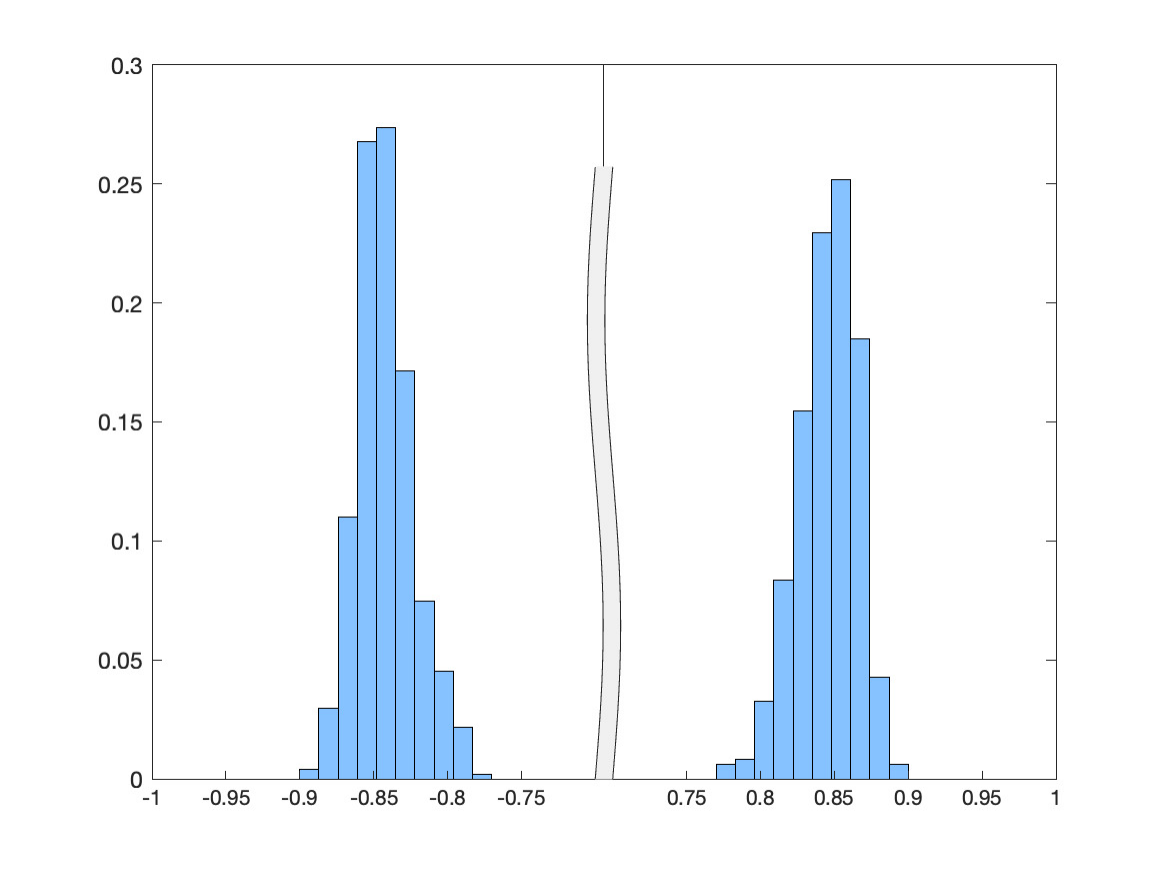} 
        \caption{Histogram of $\hat{\bu}_1^{\top}\bu_1$(spiked model)}
    \end{subfigure}
    \begin{subfigure}{0.4\textwidth}
        \centering
        \begin{tikzpicture}
            \begin{axis}[view={30}{30}, axis lines=none, width=1.2\textwidth, height= 1.2\textwidth, colormap={mycolormap}{ rgb255(0cm)=(82,124,180);   rgb255(1cm)=(127,169,205);  rgb255(2cm)=(209,226,239);}, samples=40, domain=0:360, y domain=0:0.8, zmin=0, zmax=1]
                \addplot3[surf, shader=interp,  z buffer=sort]({y*cos(x)}, {y*sin(x)}, {y});
                \addplot3[ -stealth, thick, black, domain=0.51:1 ](0, 0, {x});
                \addplot3[-stealth, thick, black,domain=0:0.8]({0.6*x}, {0}, {0.8*x});
                \node[anchor=south] at (axis cs:0,0,1) {$\bu_1$};
                \node[anchor=east] at (axis cs:0.8,0,0.7) {$\hat{\bu}_1$};
            \end{axis}
        \end{tikzpicture}
        \caption{$\hat\bu_1$ under the spiked model}
    \end{subfigure}
    \caption{
      Histogram of 
    $\hat\bu_1^{\top}\bu_1$ (a) and distribution of $\hat\bu_1$ (b)
   under the non-spiked model where $\boldsymbol{\Sigma}=\left(0.5^{|i-j|}\right)_{p\times p}$.
Histogram of 
$\hat\bu_1^{\top}\bu_1$ (c) and distribution of $\hat\bu_1$ (d)
under the spiked model where $\boldsymbol{\Sigma}_{p\times p}=\left(0.5^{|i-j|}\right)+(\sqrt{2}, \sqrt{2}, 0, \ldots, 0)^{\top}(\sqrt{2}, \sqrt{2}, 0, \ldots, 0)$ and $K=1$.
Here $p=800$, $N=1000$ and the replication is 1000.
}
    \label{fig:pca-ev}
\end{figure}
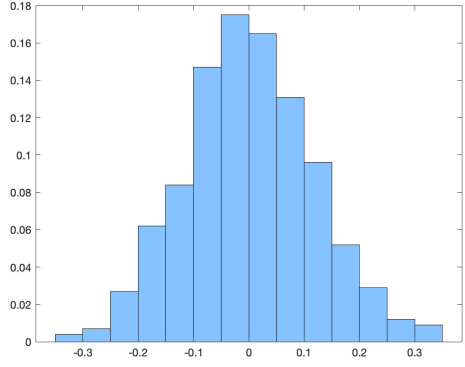
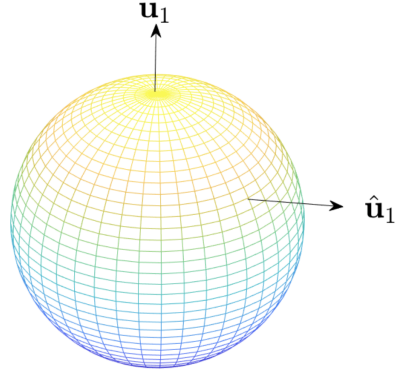
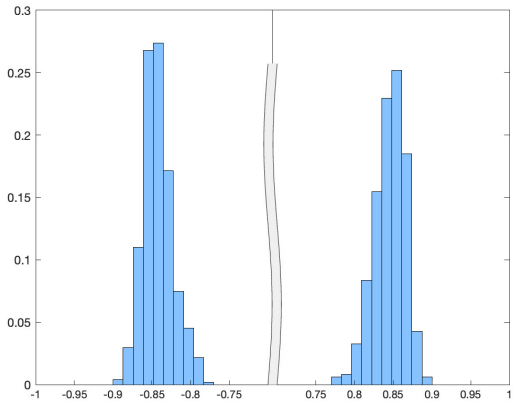
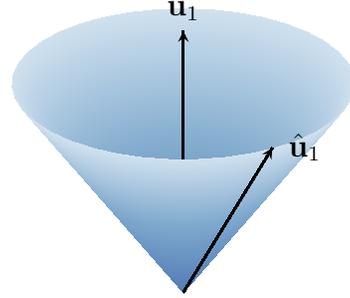

\begin{figure}[H]
     \centering
    \begin{subfigure}{0.48\textwidth}
        \centering
\includegraphics[width=\textwidth]{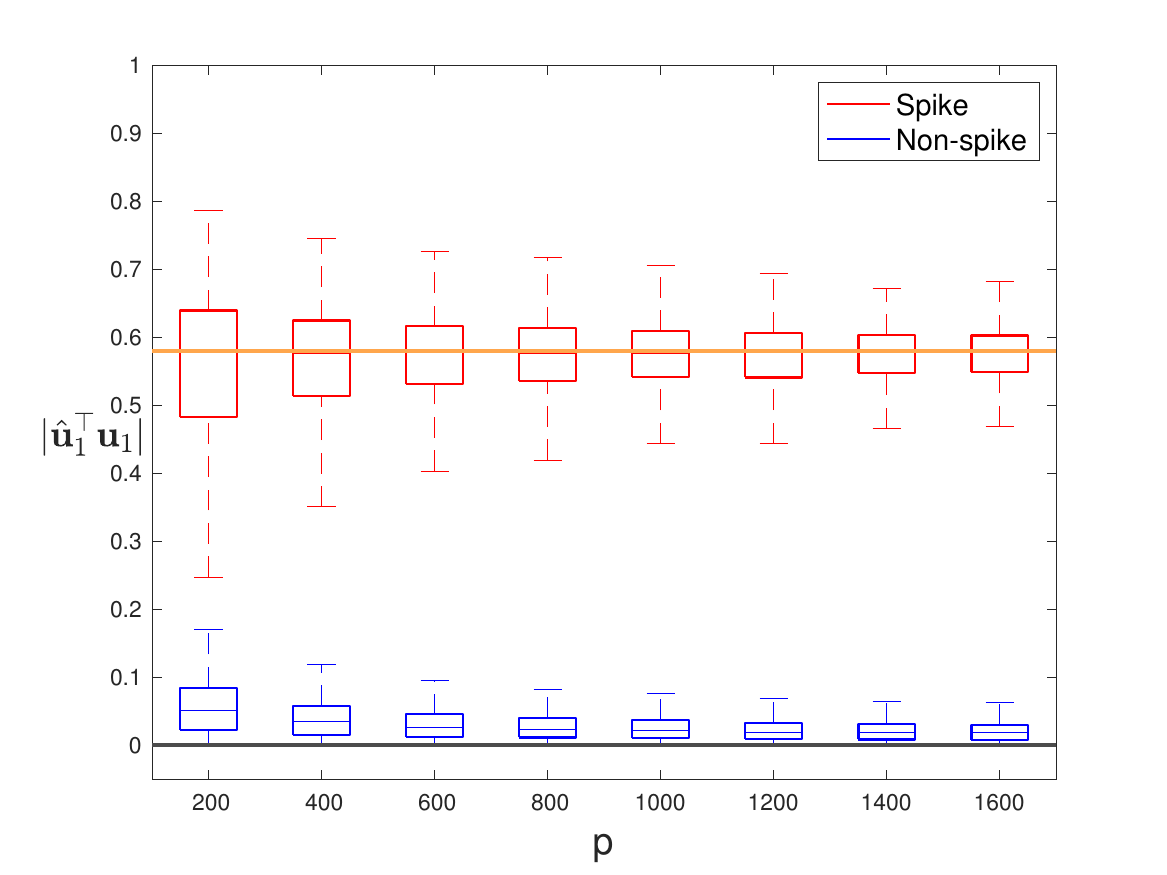}
        \caption{Box plots of $|\hat{\mathbf{u}}_1^{\top} \mathbf{u}_1|$($m=1$)}
\end{subfigure}
     \begin{subfigure}{0.48\textwidth}
        \centering
\includegraphics[width=\textwidth]{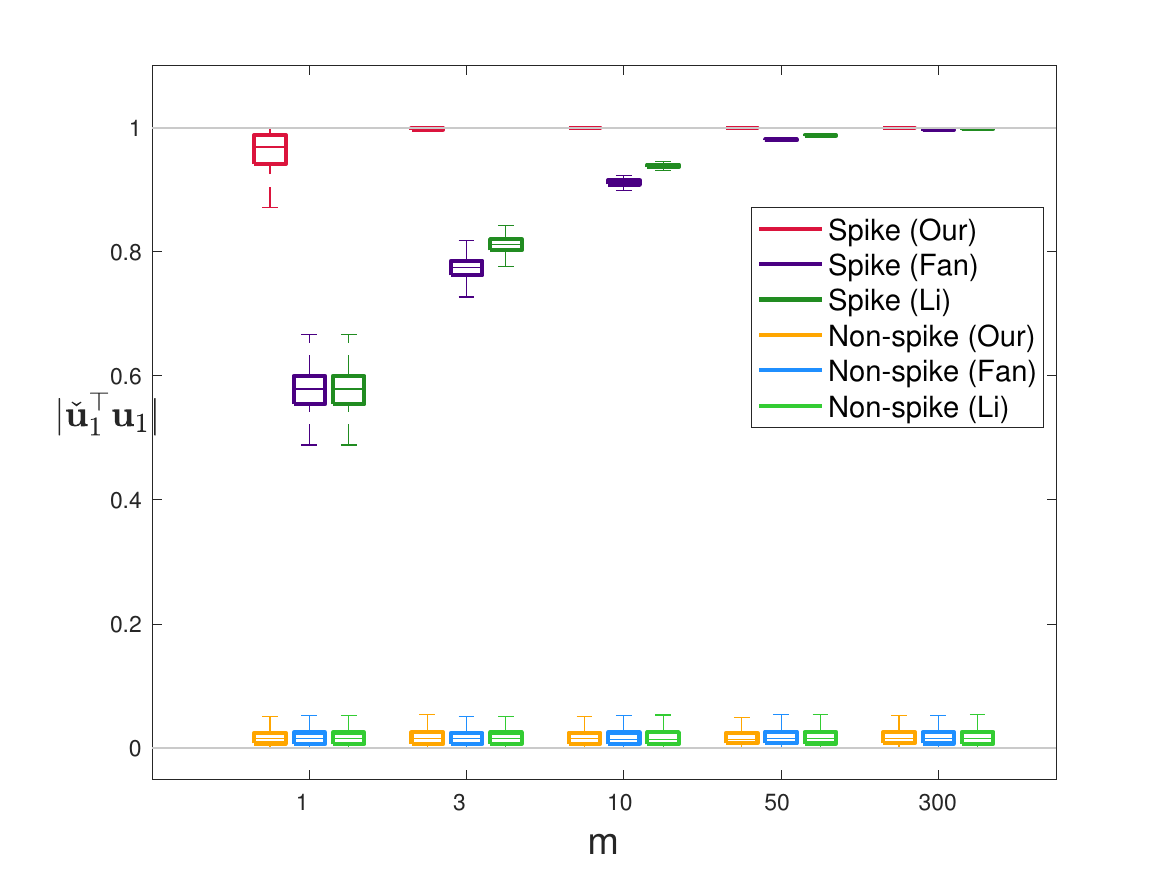}
        \caption{Box plots of $|\check{\mathbf{u}}_1^{\top} \mathbf{u}_1|$($m\geq1$)}
    \end{subfigure}
    \caption{(a) Box plots of  $|\hat{\mathbf{u}}_1^{\top} \mathbf{u}_1|$ ($m=1$) 
    under the spiked model with $K=1$,
    	$\bSigma=\operatorname{diag}\{2,1,1,\ldots,1\}$ and  the non-spiked model with $\bSigma=\bI_p$. The  orange horizontal line represents asymptotic limit of $\left|\hat{\mathbf{u}}_1^{\top} \mathbf{u}_1\right|$ under the spiked model.  The dimension $p$ ranges from $200$ to $1600$ and $p / N=0.5$. 
    	(b) Box plots of $|\check{\mathbf{u}}_1^{\top} \mathbf{u}_1|$ ($m\geq 1$) of different distributed PCA algorithms under the same model settings in (a), with $p=2000, n_{\ell}=4000,1 \leq \ell \leq m$.
    	The tuning parameter in our algorithm is set to $t=0.1$.
    	 All the box plots are based on 1000 replications. }
    \label{fig:converge}
\end{figure}

The reason we choose the spiked model is that, under high-dimensional non-spiked settings, the sample eigenvectors carry little information about the population eigenvectors. For example, when $\boldsymbol{\Sigma} = \mathbf{I}_p$, the first $K$ eigenvectors of the sample covariance matrix $\hat{\mathbf{u}}_i$ ($1 \leq i \leq K$) are asymptotically uniformly distributed on the unit sphere, and the eigenvector matrix 
nearly follows a Haar distribution, as shown in \cite{bai07}. In such cases, $\hat{\mathbf{u}}_i$ is entirely random and bears no relation to the population eigenvector $\mathbf{u}_i$.
For more general non-spiked  $\bSigma$, the sample eigenvector exhibits similar behavior, see Figure \ref {fig:pca-ev} (a)-(b).
In contrast, under the spiked model, the angle between the sample eigenvector $\hat{\mathbf{u}}_i$ and the population eigenvector $\mathbf{u}_i$ converges to a nonzero constant  as established in \cite{MR2589886,Paul2007,couillet2022random,yinpca} and shown in Figure \ref {fig:pca-ev} (c)-(d) and Figure \ref{fig:converge} 
 (a). This property allows us to effectively aggregate multiple realizations $\left\{\hat{\mathbf{u}}_i^{(\ell)}\right\}_{\ell=1}^m$ to recover the population eigenvector $\mathbf{u}_i$.


In the distributed setting, under the non-spiked model, we observe that the estimated eigenvectors produced by existing averaging-based methods (e.g., \cite{fan,li2025two}) are nearly orthogonal to the  population eigenvectors. Moreover, this issue persists even as the number of machines $m \to \infty$, with no notable improvement in estimation accuracy (see Figure~\ref{fig:converge} (b)). 
In contrast, under the spiked model, these methods exhibit markedly different behavior: the bias in the estimated eigenvectors decreases as $m$ increases, and consistency is achieved  as $m \to \infty$ (Figure~\ref{fig:converge} (b)). Consequently, with a sufficiently large number of machines $m$, the population eigenvectors can be accurately recovered from their sample counterparts.
Although existing methods fail to produce accurate estimates when $m$ is small, we show that the bias can be corrected locally on each machine by leveraging random matrix theory before aggregation. In fact, we propose a bias-corrected method based on the spiked model, 
which enables more accurate recovery even with a small number of machines, as demonstrated in Figure~\ref{fig:converge} (b).

\subsection{Bias of eigenvectors}
In this section, we present the relationship between the sample eigenvectors  and the population eigenvectors of the covariance matrix under the spiked model, which serves as the foundation for our bias-corrected method.
Specifically, \cite{yinpca} showed that under the spiked model, as $n_{\ell} \to \infty$, $p \to \infty$, $p / n_{\ell} \to c_{\ell} \in (0, \infty)$, the sample covariance eigenvector $\hat{\mathbf{u}}_i^{(\ell)}$ and the corresponding population eigenvector $\bu_i$ satisfy
\begin{align}\label{recover-space}
\left(\theta_i^{(\ell)}\right)^{-2}\be_a^{\top}\hat\bu_i^{(\ell)}\hat\bu_i^{(\ell)\top}\be_b=\be_a^{\top}\bu_i\bu_i^{\top}\be_b+o_p(1),\quad 
1\leq i\leq K,~ 1\leq \ell\leq m,~
1\leq a,b\leq p,
\end{align}
where  $\left\{\mathbf{e}_1, \ldots, \mathbf{e}_p\right\}$ denote the canonical base vectors of $\mathbb{R}^p$
and
\begin{align}\label{theta-step4}
\theta_i^{(\ell)}= \sqrt{-\frac{\varphi_i^{(\ell)}}{\hat{\lambda}_i^{(\ell)} \phi_i^{(\ell)}}},
\end{align}
with\begin{align*}
\varphi_i^{(\ell)}&=-\frac{1-(p-K) / n_{\ell}}{\hat{\lambda}_i^{(\ell)}}+\frac{1}{n_{\ell}} \sum_{j=K+1}^p \frac{1}{\hat{\lambda}_j^{(\ell)}-\hat{\lambda}_i^{(\ell)}},\\
\phi_i^{(\ell)}&=\frac{1-(p-K) / n_{\ell}}{\left(\hat{\lambda}_i^{(\ell)}\right)^2}+\frac{1}{n_{\ell}} \sum_{j=K+1}^p \frac{1}{\left(\hat{\lambda}_j^{(\ell)}-\hat{\lambda}_i^{(\ell)}\right)^2}.
\end{align*}
 Based on this result, we can reveal the relationship among the entries of $\hat{\mathbf{u}}_i^{(\ell)}=(\hat{u}_{ij}^{(\ell)})$ and   $\bu_i=(u_{ij})$, which is \begin{align}\label{recover-split-term}
\underbrace{|\hat{u}_{ij}^{(\ell)}/\theta_i^{(\ell)}|}_{\text{sample term}}=\underbrace{|u_{ij}|}_{\text{signal term}}
+\underbrace{o_p(1)}_{\text{error term}},\quad 
1\leq i\leq K,\quad 1\leq \ell\leq m,\quad 1\leq j\leq p.
\end{align}
 It's important to note that when $u_{ij}\asymp 1$ , $|u_{ij}|$ can be accurately recovered using the sample term $\left|\hat{u}_{i j}^{(\ell)} /\theta_i^{(\ell)}\right|$. However, when  $u_{ij}=o(1)$, the signal is buried in the noise, making it practically unidentifiable.

The core idea of our method is to first identify the indices corresponding to strong signals (i.e., entries where $u_{ij} \asymp 1$). We then adjust each local estimator $\hat{\bu}_{i}^{(\ell)}$ based on equation~\eqref{recover-split-term} to obtain a more accurate estimate. Finally, the corrected local estimators are aggregated to produce a global estimator. The detailed algorithm is presented in the next section.

\newpage

\section{Algorithm.}\label{sec:3}
In this section, we introduce a bias-corrected distributed PCA algorithm for estimating the matrix of the top $K$ eigenvectors $\bU_K = (\bu_1, \ldots, \bu_K)$. The proposed algorithm consists of three main stages:

\begin{itemize}
\item[] \textbf{Stage 1: Initialization.}
Each machine computes a local estimator.
\item[] \textbf{Stage 2: Signal Identification.}  
Identify strong signal components ($\bu_{ij} \asymp 1$) that allow for substantial bias correction.  
\item[] \textbf{Stage 3: Signal Recovery.}  
Apply different recovery strategies for strong signals ($\bu_{ij} \asymp 1$) and weak signals ($\bu_{ij} = o(1)$), respectively.  
\end{itemize}

We adopt the standard notation $[q] \triangleq \{1,\ldots,q\}$ ($q\in \mathbb N$) for index sets.

\subsection{Initialization.}

In the initialization stage, each machine independently computes a local estimator.
Specifically, for machine $\ell \in[m]$, we compute the sample covariance matrix and obtain its eigenvalues $\left\{\hat{\lambda}_i^{(\ell)}: i \in[p]\right\}$ and the top $K$ eigenvectors $\left\{\hat{\mathbf{u}}_i^{(\ell)}: i \in[K]\right\}$.
Subsequently, we derive the eigenvalue-dependent correction factors $\left\{\theta_i^{(\ell)}: i \in[K]\right\}$ based on the local eigenvalues.

This initialization procedure is formalized in Algorithm 1-1.

\begin{algorithm}[H]
\begin{algorithmic}
\caption{ \textbf{1-1: Initialization} } 
\smallskip
\STATE 1. On each machine \(\ell \in [m]\), compute the local sample covariance matrix \(\bS_{n_\ell}^{(\ell)} = \bX_{n_\ell}^{(\ell)} \bX_{n_\ell}^{(\ell)\top} / n_\ell\), where \(\bX_{n_\ell}^{(\ell)} \in \mathbb{R}^{p \times n_\ell}\) is the locally stored data matrix. 
From  \(\bS_{n_\ell}^{(\ell)}\), extract its eigenvalues \(\{\hat{\lambda}_i^{(\ell)}: i \in [p]\}\) and the top \(K\) eigenvectors \(\{\hat{\mathbf{u}}_i^{(\ell)}: i \in [K]\}\), and finally compute the eigenvalue-dependent correction factor \(\theta_i^{(\ell)}\) defined in \eqref{theta-step4}.  
\end{algorithmic}
\end{algorithm}

\subsection{Identification of signals.}
\begin{figure}[h]
\centering
\begin{tikzpicture}
		\draw[thick] (0, 1) -- (6, 1); 
		\node[left] at (0, 1) {$\bu_1$}; 
		\draw[Blue4, thick] (0, 0.8) rectangle (1, 1.2); 
		\draw[Red2, thick] (1.05, 0.8) rectangle (6, 1.2); 
		\node[below] at (3.5, 0.9) {$\mathcal{A}_1^c(t)$}; 
		\node[below] at (0.5, 0.9) {$\mathcal{A}_1(t)$}; 
		
		\draw[thick] (0, 0) -- (6, 0); 
		\node[left] at (0, 0) {$\bu_2$}; 
		\draw[Blue4, thick] (0.8, -0.2) rectangle (1.8, 0.2); 
		\draw[Red2, thick] (0, -0.2) rectangle (0.75, 0.2); 
		\draw[Red2, thick] (1.85, -0.2) rectangle (6, 0.2); 
		\node[below] at (3.9, -0.1) {$\mathcal{A}_2^c(t)$}; 
		\node[below] at (0.36, -0.1) {$\mathcal{A}_2^c(t)$}; 
		\node[below] at (1.3, -0.1) {$\mathcal{A}_2(t)$}; 
		
		\node[below] at (3, -0.8) {$...$ }; 
		
		\draw[thick] (0, -1.6) -- (6, -1.6); 
		\node[left] at (0, -1.6) {$\bu_k$}; 
		\draw[Blue4, thick] (1, -1.8) rectangle (2, -1.4); 
		\draw[Red2, thick] (0, -1.8) rectangle (0.95, -1.4); 
		\draw[Red2, thick] (2.05, -1.8) rectangle (6, -1.4); 
		\node[below] at (4, -1.7) {$\mathcal{A}_k^c(t)$}; 
		\node[below] at (0.5, -1.7) {$\mathcal{A}_k^c(t)$}; 
		\node[below] at (1.5, -1.7) {$\mathcal{A}_k(t)$}; 

	\draw[Blue5, thick] (-0.03, -2.5) rectangle (2.03, 1.4); 
\draw[Red1, thick] (2.05, -2.5) rectangle (6.03, 1.4);  
\node[below] at (1, -2.6) {$\mathcal{A}(t)$}; 
\node[below] at (3.5, -2.6) {$\mathcal{A}^c(t)$}; 
\end{tikzpicture}
\caption{Diagram of the composite recovery region \(\mathcal{A}(t)\) and its complement \(\mathcal{A}^c(t)\).}
\label{A and A^c}
\end{figure}
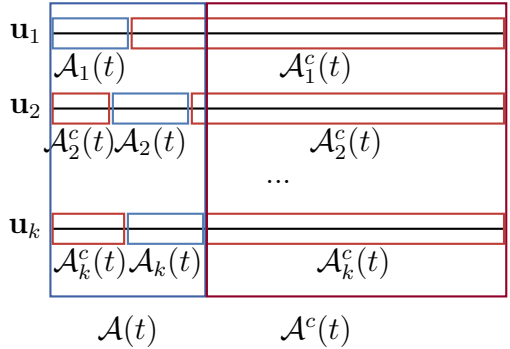

This stage detects coordinates with strong signals ($u_{ij} \asymp 1$) that allow for a substantial bias correction. For each eigenvector \(\bu_i\) (\(i \in [K]\)), we define the strong signal index set as
$$
\mathcal{A}_i = \{j \in [p] : u_{ij} \asymp 1\},\quad i\in [K].
$$
Since the index sets 
\(\{\mathcal{A}_i\}_{i=1}^K\) 
may vary significantly across the eigenvectors, we define a composite recovery region as  
\[
\mathcal{A} = \bigcup_{i=1}^K \mathcal{A}_i.  
\]  
In practical implementation, we employ a threshold-approximate $$ \mathcal{A} _i (t) = \{j \in [p] : |u_{ij}| > t\},\ \mathcal{A}(t)= \bigcup_{i=1}^K \mathcal{A}_i(t).
$$
where the parameter \(t > 0\) is a tuning parameter that controls the level of signal inclusion. 
This unified set \(\mathcal{A}(t)\) ensures the inclusion of all strong signal coordinates while preserving computational efficiency. Its complement  
$\mathcal{A}^c(t) = [p]\setminus \mathcal{A}(t)$, see Figure \ref{A and A^c} for illustration.

To estimate $\mathcal{A}(t)$, the procedure operates as follows:
\begin{enumerate}
    \item \textbf{Local Computation} (per machine $\ell$):  
   For each $i\in [K]$, compute threshold indicators:
    \[
    \hat I_{ij}^{(\ell)} = \mathbbm{1}\left(|\hat{u}_{ij}^{(\ell)}/\theta_i^{(\ell)}| > t\right),
    \quad j \in [p],\]
    where $\hat{u}_{ij}^{(\ell)}$ is the $j$-th element of the local sample eigenvector $\hat{\bu}_i^{(\ell)}$.

    \item \textbf{Global Aggregation} (central processor):  
    Construct the final estimate by majority voting:
    \[
    \widehat{\mathcal{A}}(t) = \bigcup_{i=1}^K \widehat{\mathcal{A}}_i(t),\ 
    \widehat{\mathcal{A}}_i(t) = \left\{j \in [p] : \frac{1}{m}\sum_{\ell=1}^m \hat I_{ij}^{(\ell)} > \frac{1}{2}\right\}.
    \]
\end{enumerate}

This signal identification procedure is specified in Algorithm 1-2.

\begin{algorithm}[H]
\begin{algorithmic}
\caption{ \textbf{1-2: Identification of signals} } 

\STATE 1. For each machine $\ell\in [m]$, compute threshold indicators
    $$
    \hat I_{ij}^{(\ell)} = \mathbbm{1}\left(|\hat{u}_{ij}^{(\ell)}/\theta_i^{(\ell)}| > t\right),
    \quad i\in [K], ~ j \in [p],
    $$ 
    where $\hat u_{ij}^{(\ell)}$ is the $j$-th element of $\hat\bu_i^{(\ell)}$ and
$\theta_i^{(\ell)}$ is the bias factor from \eqref{theta-step4}. 

\STATE 2. Estimate the index set $\mathcal{A}(t)$ by 
 \[
    \widehat{\mathcal{A}}(t) = \bigcup_{i=1}^K \widehat{\mathcal{A}}_i(t),\ 
    \widehat{\mathcal{A}}_i(t) = \left\{j \in [p] : \frac{1}{m}\sum_{\ell=1}^m \hat I_{ij}^{(\ell)} > \frac{1}{2}\right\}.
    \]
\end{algorithmic}
\end{algorithm}

\subsection{Signal recovery.}
In this stage, we construct a global  estimator  $\check{\bU}_K$ for the eigenmatrix 
 \(\mathbf{U}_K\). Given the  estimate $\widehat{\mathcal{A}}(t)$ above, if $\widehat{\mathcal{A}}(t)=\emptyset$, we estimate the entries of \(\mathbf{U}_K\) using a unified procedure. Specifically, for each \( i \in [K] \), compute the aggregated matrix
\begin{align}\label{nex-Si-tilde}
   \widetilde{\mathbf{S}}_i = \frac{1}{m}\sum_{\ell=1}^m \left[\theta_{i}^{(\ell)}\right]^{-2} \hat{\mathbf{u}}_i^{(\ell)} \hat{\mathbf{u}}_i^{(\ell)\top},
\end{align}
where each summand \([\theta_{i}^{(\ell)}]^{-2}\hat{\mathbf{u}}_i^{(\ell)} \hat{\mathbf{u}}_i^{(\ell)\top}\) represents a rank-$1$ bias-corrected estimate of $\bu_i\bu_i^\top$. Then we perform an eigendecomposition on \(\widetilde{\mathbf{S}}_i\) and take its principal eigenvector \(\tilde{\mathbf{u}}_i\) (corresponding to the largest eigenvalue) as our refined estimate of $\bu_i$.
The complete estimate of $\bU_K$ is given by
$\check{\mathbf{U}}_{K}=\tilde{\mathbf{U}}_{K}\left(\tilde{\mathbf{U}}_{K}^{\top} \tilde{\mathbf{U}}_{K}\right)^{-1 / 2}$, where
\(
\tilde{\mathbf{U}}_K = \left(\tilde{\mathbf{u}}_1, \ldots, \tilde{\mathbf{u}}_K\right).
\)

If  $\widehat{\mathcal{A}}(t)\neq\emptyset$, we apply different recovery strategies for strong signals $\left(\mathbf{u}_{i j} \asymp 1\right)$ and weak signals $\left(\mathbf{u}_{i j}=o(1)\right)$ respectively and construct an estimator for the eigenmatrix 
 \(\mathbf{U}_K\). Denote the cardinality of $\widehat{\mathcal{A}}(t)$ as \( K_t \). 
We partition each eigenvector $\bu_i$ into two sub-vectors:
$$
\bu_{i1} = \bu_i[\widehat{\mathcal{A}}(t)] \in \mathbb{R}^{K_t}  
\quad \text{and} \quad  
\bu_{i2} = \bu_i[\widehat{\mathcal{A}}^c(t)] \in \mathbb{R}^{p-K_t}, \quad i \in [K].  
$$  
These two components will be estimated separately through similar partitions of $\tilde\bu_{i}$ and $\{\hat\bu_{i}^{(\ell)},\ell\in[m]\}$.
Specifically, we estimate strong signals $\bu_{i1}$ as
\begin{align*}
\bar\bu_{i1}=\arg\min_{\bu\in\mathbb R^{K_t}, ~||\bu||\leq 1}
 \left\|
 \frac{1}{m}\sum_{\ell=1}^m\frac{\hat\bu_{i1}^{(\ell)}\hat\bu_{i1}^{(\ell)\top}}{[\theta_i^{(\ell)}]^2}-\bu\bu^\top
 \right\|_F,
\end{align*}
 where
$\bar\bu_{i1}$ only depends on local estimators $
\hat\bu_{i1}^{(\ell)} = \hat\bu_i^{(\ell)}[\widehat{\mathcal{A}}(t)]$. As for weak signals $\bu_{i2}$, we estimate it using 
$\bar\bu_{i1}$ and 
$\widetilde\bS_i$ defined in \eqref{nex-Si-tilde}, as follows:
\begin{align*}
\bar\bu_{i2}=\frac{\sign(\bar\bu_{i1}^\top\tilde\bu_{i1})}{||\tilde\bu_{i2}||/(1-||\bar\bu_{i1}||^2)^{1/2}}\cdot\tilde\bu_{i2},
\end{align*}
where
$\tilde\bu_{i1} = \tilde\bu_i[\widehat{\mathcal{A}}(t)], 
\tilde\bu_{i2} = \tilde\bu_i[\widehat{\mathcal{A}}^c(t)] $. 
The factor in the denominator of $\bar\bu_{i2}$ is chosen such that $||\bar\bu_{i1}||^2+||\bar\bu_{i2}||^2=1$.

To further refine the intermediate estimate $\left\{\overline{\mathbf{u}}_1, \ldots, \overline{\mathbf{u}}_K\right\}$, we incorporate both sparsity and orthogonality constraints inherent in PCA. The resulting refined estimate  $\check{\mathbf{U}}_K = (\check\bu_1, \ldots, \check\bu_K)$ is obtained through the following procedure.
\begin{itemize}
    \item[1.] \textbf{Detection of sparse vectors.} 
For each eigenvector $\bu_i$, $i\in[K]$,  we assess its sparsity using the following criterion:
\begin{align}\label{spares-condition}
\bar\bu_{i1}^\top\bar\bu_{i1} > 1 - 2m^{-1/4}p^{-1/2}.
\end{align}
If this condition is satisfied, we construct a sparse estimator $\check\bu_i$ as
$$\check\bu_{i}[\widehat{\mathcal A}(t)] = \bar\bu_{i1}/\|\bar\bu_{i1}\|,\ \check\bu_{i}[\widehat{\mathcal A}^c(t)] = \mathbf{0}_{p-K_t}.$$

\item[2.] \textbf{Orthogonal projection for non-sparse eigenvectors.} 
For clarity, we assume that $K_0$ out of the $K$ estimated eigenvectors $\left\{\overline{\mathbf{u}}_1, \ldots, \overline{\mathbf{u}}_K\right\}$ are non-sparse. We collect these vectors into the matrix $\overline{\mathbf{U}}_{K_0}$ and perform orthogonalization via
$
\check{\mathbf{U}}_{K_0}=\overline{\mathbf{U}}_{K_0}\left(\overline{\mathbf{U}}_{K_0}^{\top} \overline{\mathbf{U}}_{K_0}\right)^{-1 / 2}.
$
This guarantees that the resulting non-sparse eigenvectors are orthogonal, i.e., $\check{\mathbf{U}}_{K_0}^{\top} \check{\mathbf{U}}_{K_0}=\mathbf{I}_{K_0}$.
\end{itemize}
This recovery process is formalized in Algorithm 1–3.

\begin{algorithm}[h!]
\begin{algorithmic}
\caption{ \textbf{1-3: Signal recovery --- construct $\check{\bU}_K=\left(\check{\bu}_1, \cdots, \check{\bu}_K\right)$ to estimate $\bU_K$}} 
\IF{$\widehat{\mathcal{A}}(t)=\emptyset$}
\STATE $\check{\mathbf{U}}_K=\tilde{\mathbf{U}}_K\left(\tilde{\mathbf{U}}_K^{\top} \tilde{\mathbf{U}}_K\right)^{-1 / 2}$, 
where $\tilde{\mathbf{U}}_K=\left(\tilde{\mathbf{u}}_1, \ldots, \tilde{\mathbf{u}}_K\right)$ and $\tilde{\mathbf{u}}_i$ is the principal eigenvector (corresponding to the largest eigenvalue) of 
$
\widetilde{\mathbf{S}}_i=\frac{1}{m} \sum_{\ell=1}^m\left[\theta_i^{(\ell)}\right]^{-2} \hat{\mathbf{u}}_i^{(\ell)} \hat{\mathbf{u}}_i^{(\ell) \top},\ i \in[K].
$ 
\ELSIF{$\widehat{\mathcal{A}}(t)\neq\emptyset$}
\STATE \texttt{//Generate intermediate estimate}\\
 compute the strong signals estimates $\bar\bu_{i1}$ 
 and  weak signals estimates $\bar\bu_{i2}$
 as
\begin{align*}
\bar\bu_{i1}=\arg\min_{\bu\in\mathbb R^{K_t}, ~||\bu||\leq 1}
 \left\|
 \frac{1}{m}\sum_{\ell=1}^m\frac{\hat\bu_{i1}^{(\ell)}\hat\bu_{i1}^{(\ell)\top}}{[\theta_i^{(\ell)}]^2}-\bu\bu^\top
 \right\|_F,\ 
\bar\bu_{i2}=\frac{\sign(\bar\bu_{i1}^\top\tilde\bu_{i1})}{||\tilde\bu_{i2}||/(1-||\bar\bu_{i1}||^2)^{1/2}}\cdot\tilde\bu_{i2},
\end{align*}
where $\hat\bu_{i1}^{(\ell)} = \hat\bu_i^{(\ell)}[\widehat{\mathcal{A}}(t)]$ and $\tilde\bu_{i2} = \tilde\bu_i[\widehat{\mathcal{A}}^c(t)].$\\
\texttt{//Incorporate sparsity and orthogonality}
\IF{$\|\bar{\bu}_{i1}\|^2 \geq 1 - \frac{2}{m^{1/4}p^{1/2}}$}
\STATE $
\check{\bu}_i[\widehat{\mathcal{A}}(t)] = \bar{\bu}_{i1} / \|\bar{\bu}_{i1}\|, \ \check{\bu}_i[\widehat{\mathcal{A}}^c(t)] = \mathbf{0}_{p-K_t}.  
  $  
\ELSIF{$\|\bar{\bu}_{i1}\|^2<1 - \frac{2}{m^{1/4}p^{1/2}}$}
\STATE we incorporate this vector $\bar\bu_i=\left(\bar\bu_{i1}^{\top},\bar\bu_{i2}^{\top}\right)^{\top}$
into the matrix $\overline{\mathbf{U}}_{K_0}$.
\ENDIF\\
Perform orthogonalization via
$
\check{\mathbf{U}}_{K_0}=\overline{\mathbf{U}}_{K_0}\left(\overline{\mathbf{U}}_{K_0}^{\top} \overline{\mathbf{U}}_{K_0}\right)^{-1 / 2}.
$
\ENDIF
\end{algorithmic}
\end{algorithm}

\begin{algorithm}[h]
\begin{algorithmic}\label{algorithm1}
\caption{ \textbf{1: Cov-PCA} } 
\STATE 1. Initialization: Algorithm 1-1.
\STATE 2. Signal identification: Algorithm 1-2.
\STATE 3. Signal recovery: Algorithm 1-3.
\STATE Output: $\check{\bU}_K$.
\end{algorithmic}
\end{algorithm}

\noindent
 We present the complete procedure of our distributed PCA algorithm in \textbf{Algorithm 1}. This method can be further extended to estimate the top $K$-dimensional eigenspace of the correlation matrix
$
\bR = \mathbf{D}^{-\frac{1}{2}} \boldsymbol{\Sigma} \mathbf{D}^{-\frac{1}{2}},
$
where $\mathbf{D} = \operatorname{diag}(\boldsymbol{\Sigma})$. In this extension, the sample covariance matrix is replaced by the sample correlation matrix accordingly.

\section{Asymptotic properties.}\label{sec:4}
\subsection{Sample average of projection matrices.}
In this section, we present the asymptotic properties of the empirical average of projection matrices  
\[  
\frac{1}{m} \sum_{\ell=1}^m \hat{\mathbf{u}}_i^{(\ell)} \hat{\mathbf{u}}_i^{(\ell)\top}, \quad i \in [K],  
\]  
as the number of local machines \( m \) tends to infinity. Here, each \( \hat{\mathbf{u}}_i^{(\ell)} \) denotes the \( i \)-th sample eigenvector computed on the \( \ell \)-th machine. This result plays a fundamental role in establishing the consistency of our distributed PCA estimator. 
The properties we derive are  more refined than those in existing literature.  Furthermore, our proof strategy is fundamentally different, as it adopts a novel approach grounded in random matrix theory.  
Specifically, by employing two key tools—the resolvent matrix of sample covariance matrices and contour integration—we decompose the empirical projection matrix $\hat\bu_i^{(\ell)}\hat\bu_i^{(\ell)\top}$ into a signal component and a residual component. We then show that the aggregation process removes the residual component, enabling us to derive the limiting behavior of the averaged empirical projection matrices $\hat\bu_i^{(\ell)}\hat\bu_i^{(\ell)\top}$.
Most importantly, our theoretical guarantees are established under significantly weaker conditions—we only assume that the data distribution has a finite sixth moment, whereas prevailing works typically require sub-Gaussianity and symmetric innovation assumptions. Specifically, the theoretical analysis of the empirical average of projection matrices is built upon the following assumptions.





\begin{assumption}\label{as-moment}
The sample is $\{\bx_i=\bSigma^{\frac{1}{2}}\bw_i,~ i=1,\ldots, N\}$, where $(\bw_1,\dots,\bw_N)=(w_{ij})_{p\times N}$ consists of i.i.d.\ random variables 
satisfying 
$$
\mathbb{E}\left(w_{i j}\right)=0, \quad \mathbb{E}\left(w_{i j}^2\right)=1, \quad \mathbb{E}\left(w_{i j}^6\right)<\infty .
$$
\end{assumption}

\begin{assumption}\label{as-mp}
The dimension $p$ and the sample sizes $\left\{n_1, \ldots, n_m\right\}$, $\sum_{\ell=1}^{m}n_{\ell}=N$, all tend to infinity such that
$$
c_{n_\ell}\triangleq \frac{p}{n_{\ell}}\rightarrow c_{\ell}\quad\text{and}\quad \max_{1\leq \ell\leq m} \{c_{n_\ell}\}< c_0,
$$
where $\{c_{\ell}\}$ are positive constants bounded by $c_0<\infty$.
The number of servers is $m=O(p)$. 
\end{assumption}
\begin{assumption}\label{as-spike}
The empirical spectral distribution  $H_p$ of $\bSigma$ converges weakly to a probability distribution $H$ with compact support $\Gamma_H \subset \mathbb{R}^+$. The $K$ largest eigenvalues of $\bSigma$, $\lambda_1 > \lambda_2 > \dots > \lambda_K$, are spike eigenvalues located outside $\Gamma_H$. They satisfy the following conditions: 
\begin{align}\label{distance-spikes}
\min_{1 \leq k \leq K} (\lambda_k - \lambda_{k+1}) > d_0, \quad \psi_{c_0, H}'(\lambda_K) \triangleq 1 - c_0 \int \frac{t^2}{(x - \lambda_K)^2} d H(t)> d_0,  
\end{align}  
for some constant $d_0>0$.
For correlation matrix, we impose an analogous assumption in which the covariance matrix $\bSigma$ is replaced by the correlation matrix $\bR$.
\end{assumption}

\begin{remark}
Assumptions \ref{as-moment} and \ref{as-mp} define a high-dimensional framework for the observed data \((\bx_i)\), where the population maintains a linear dependence structure among its components with only finite sixth moments. The population dimension \(p\) and the sample sizes \(\{n_\ell\}\) are allowed to grow proportionally to infinity, while the number of servers \(m\) can be finite or diverge with \(p\). These assumptions form the theoretical foundation for our random matrix analysis. 
\end{remark}

\begin{remark}
Assumption \ref{as-spike} specifies a spiked population model for the covariance matrix $\bSigma$. See \cite{Johnstone01} and \cite{Bai2012}. The conditions in \eqref{distance-spikes} guarantee that each server's sample covariance matrix will exhibit exactly $K$ leading eigenvalues that separate from the bulk spectrum. These spikes and their associated eigenvectors provide the basis for reconstructing the principal eigenspace of the population covariance matrix.
\end{remark}


\begin{theorem}\label{th-main}
Suppose that Assumptions \ref{as-moment}-\ref{as-spike} hold.
Then the Frobenius norm
\begin{align*}
&\left \|\frac1m\sum_{\ell=1}^m\hat\bu_i^{(\ell)}\hat\bu_i^{(\ell)\top}
-\frac1m\sum_{\ell=1}^m\E\left(\bu_i^\top\hat\bu_i^{(\ell)}\right)^2\bu_i\bu_i^\top
\right \|_F\xrightarrow{i.p.}0,\quad i\in [K],
\end{align*}
as $m\to\infty$. 
\end{theorem}

Theorem \ref{th-main} shows that, in terms of the Frobenius norm, the first principal eigenvector of the averaged projection matrix $(1/m)\sum_{\ell=1}^m\hat\bu_i^{(\ell)}\hat\bu_i^{(\ell)\top}$ can consistently estimate $\bu_i$ as $m \to \infty$. Its proof relies on a key decomposition of each projection matrix:  
\begin{align}\label{paper-residual}
\hat\bu_i^{(\ell)}\hat\bu_i^{(\ell)\top} = \left(\bu_i^\top\hat\bu_i^{(\ell)}\right)^2 \bu_i\bu_i^\top + 
{\mathcal R}_{n_\ell}^{(\ell)},  
\end{align}
where the first term represents a biased estimate of $\bu_i\bu_i^\top$ with the bias governed by the inner product $\bu_i^\top\hat\bu_i^{(\ell)}$ and the second term, $ {\mathcal R}_{n_\ell}^{(\ell)}$, stands for residual error. In high-dimensional settings, neither of the two terms vanishes asymptotically. 
In particular, the absolute value of the inner product $|\bu_i^\top\hat\bu_i^{(\ell)}|$ converges to a constant strictly less than $1$ \citep{MR2782201}, preventing the first term from becoming unbiased. Accordingly, the residual error $\mathcal R_{n_\ell}^{(\ell)}$ remains non-negligible, as evidenced by the unit norm constraints $||\hat\bu_i^{(\ell)}||=||\bu_i||=1$. 
The key insight is that while individual residuals remain significant, their ensemble average converges to zero as $m\to\infty$, thus guaranteeing convergence of the eigenvector estimate.

\begin{remark}\label{th-fan}
As an application of Theorem \ref{th-main}, we can obtain the consistency of the eigenspace estimate proposed in the seminal work \cite{fan}. Their approach aggregates the sample eigenvectors as  
$
\frac{1}{m}\sum_{\ell=1}^m\sum_{i=1}^K\hat\bu_i^{(\ell)}\hat\bu_i^{(\ell)\top},  
$ 
and uses its $K$ principal eigenvectors to estimate $\bU_K$, referred to as $\tilde \bU_{K, Fan}$. 
\cite{fan} established a non-asymptotic bound and showed that $\left\|\tilde \bU_{K, Fan}\tilde \bU_{K, Fan}^\top -\bU_K\bU_K^\top\right\|_F \xrightarrow{i . p .} 0
$ 
as $m\to\infty$ under the assumption of sub-Gaussianity and symmetric innovations (e.g., elliptical distributions). 
Our Theorem~\ref{th-main} extends this result under significantly weaker assumptions on the data distribution. Specifically, we do not require symmetric innovations and only assume that the sixth moment is finite.
The detailed proof is provided in Section S8 in the Supplementary Material. This extension demonstrates the robustness of their estimator under more general distributions.
\end{remark}

Theorem \ref{th-main} can also be  extended to the correlation matrix setting. Let $\left\{\mathbf{v}_i\right\}_{i=1}^K$ denote the top $K$ eigenvectors of the population correlation matrix $\mathbf{R}$, and  $\left\{\hat{\mathbf{v}}_i^{(\ell)}\right\}_{i=1}^K$ be the top $K$ eigenvectors of the sample correlation matrix
$$
\mathbf{R}_n^{(\ell)}=\left[\operatorname{diag}\left(\mathbf{S}_{n_{\ell}}^{(\ell)}\right)\right]^{-1 / 2} \mathbf{S}_{n_{\ell}}^{(\ell)}\left[\operatorname{diag}\left(\mathbf{S}_{n_{\ell}}^{(\ell)}\right)\right]^{-1 / 2},\quad \ell\in[K],
$$ computed on the $\ell$-th machine. Then, we have the following result.
\begin{theorem}\label{th-cor} 
Suppose that Assumptions \ref{as-moment}-\ref{as-spike} hold.
Then the Frobenius norm
\begin{align*}
&\left \|\frac1m\sum_{\ell=1}^m\hat\bv_i^{(\ell)}\hat\bv_i^{(\ell)\top}
-\frac1m\sum_{\ell=1}^m\E\left(\bv_i^\top\hat\bv_i^{(\ell)}\right)^2\bv_i\bv_i^\top
\right \|_F\xrightarrow{i.p.}0,\quad i\in [K],
\end{align*}
as $m\to\infty$.
\end{theorem}
The proof of Theorem~\ref{th-cor} builds upon Theorem~\ref{th-main}. The key idea is to leverage the discrepancy between the sample covariance matrix and the sample correlation matrix,
$$
\left\|\mathbf{S}_{n_{\ell}}^{(\ell)}-\mathbf{R}_{n_{\ell}}^{(\ell)}\right\|,
$$
to control the deviation between their respective eigenspaces,
$$
\left\|\hat{\mathbf{v}}_i^{(\ell)} \hat{\mathbf{v}}_i^{(\ell) \top}-\hat{\mathbf{u}}_i^{(\ell)} \hat{\mathbf{u}}_i^{(\ell) \top}\right\|_F,\quad 
i \in[K] .
$$
This technique enables us to transfer the consistency guarantees from the covariance matrix setting to the correlation matrix setting. The complete proof is provided in Section S2 in the Supplementary Material.

 \subsection{Consistency of $\check \bU_K$ as $m\to\infty$.}

This section establishes the asymptotic consistency of our estimate $\check\bU_k$ as the number of machines $m \to \infty$. We assess
the estimation error through the Frobenius distance:
$$
\rho=\left\|\check\bU_K\check\bU_K^\top-\bU_K \bU_K^\top
\right\|_F.
$$
To begin with, we propose the following structural assumption on the elements of  population eigenvectors, which states that the strong signals ($\mathbf{u}_{i j} \asymp 1$) and weak signals ($\mathbf{u}_{i j}=o(1)$) are asymptotically separated.


\begin{assumption}\label{as-indentification}
There exists a threshold \( t_0\) such that
$$
\min_{j\in\mathcal A_i}|u_{ij}|>t_0 \quad \text{and}\quad 
\max_{j\notin\mathcal A_i}|u_{ij}|\to0,\quad i\in [K].
$$
\end{assumption}

\begin{theorem}\label{th-identification}
Suppose that Assumptions \ref{as-moment}-\ref{as-indentification} hold with $t_0>t$. Then we have    $$
\mathbb{P}\left(\widehat{\mathcal{A}}(t) \neq \mathcal{A}\right) \rightarrow 0.
$$
\end{theorem}
\begin{remark}
 Theorem \ref{th-identification} establishes the consistency of signal identification in our algorithm. Notably, this consistency holds regardless of the number of machines 
$m$. In other words, it remains valid whether 
$m$ is finite or $m\to \infty$.
\end{remark}
\begin{theorem}\label{th-new}
Suppose that Assumptions \ref{as-moment}-\ref{as-indentification} hold with $t_0> t$. Then the Frobenius distance $\rho=\|\check\bU_K\check\bU_K^\top-\bU_K \bU_K^\top\|_F\xrightarrow{i.p.}0$ as $m\to\infty$.
\end{theorem}

\subsection{Properties of $\check \bU_K$ for finite $m$.}
In this section, we present the theoretical properties of our estimator $\check \bU_K$ when $m$ is finite. In particular, when the population eigenvectors are sparse, our algorithm guarantees consistent estimation even for finite $m$, as formalized in the following theorem.

\begin{theorem}\label{th-sparse}
Suppose that Assumptions \ref{as-moment}-\ref{as-indentification} hold with $t_0>t$. If the eigenvectors $\{\bu_1,\ldots, \bu_K\}$ are sparse vectors such that 
$\sum_{j\in\mathcal A_i}u_{ij}^2\to 1$, $i\in [K],$
then $\rho=\|\check\bU_K\check\bU_K^\top-\bU_K \bU_K^\top\|_F\xrightarrow{i.p.} 0$, as $N\to\infty$.
\end{theorem}

A key reason why our estimator remains consistent even when $m$ is finite lies in the sparsity of the population eigenvectors. In such cases, our algorithm can accurately identify and recover all dominant signal components $\left\{u_{i j}: j \in \mathcal{A}_i, i \in[K]\right\}$. Since $\sum_{j \in \mathcal{A}_i} u_{i j}^2 \rightarrow 1$, the residual component $\mathcal{R}_{n_{\ell}}^{(\ell)}$ defined in~\eqref{paper-residual} becomes negligible even for finite $m$.

In contrast, for non-sparse eigenvectors, consistency cannot be achieved for finite $m$. The Frobenius error $\rho=\left\|\check{\mathbf{U}}_K \check{\mathbf{U}}_K^{\top}-\mathbf{U}_K \mathbf{U}_K^{\top}\right\|_F$ does not vanish because the residual component $\mathcal{R}_{n_{\ell}}^{(\ell)}$ cannot be fully eliminated through finite averaging.  However, our refinement of the strong signals can substantially reduce $\rho$. We illustrate this advantage in the special case where $m=1$.

\begin{theorem}\label{th-nonsparse}
Suppose that Assumptions \ref{as-moment}-\ref{as-indentification} hold with the number of machines $m=1$ and the threshold $t_0> t$. 
If $\mathcal A=
\cup_{i=1}^K\mathcal A_i \neq \emptyset $ and $\mathcal{A}_i \cap \mathcal{A}_j=\emptyset$ for $i\neq j\in[K]$, then 
\begin{align*}
\left\|\sum_{i=1}^K\hat\bu_i\hat\bu_i^{\top}-\bU_K\bU_k\right\|_F^2 
=\left\|\sum_{i=1}^K\check\bu_i\check\bu_i^{\top}-\bU_K\bU_k\right\|_F^2 +\mathscr{C}(\mathcal A)+o_p(1),\quad N\to\infty,
\end{align*}
where
\begin{align*}
\mathscr{C}(\mathcal A)=2\sum_{i=1}^{K_0}\left[\left(||\bu_{i1}||^2+\sqrt{\frac{1-||\bu_{i1}||^2}{1-||\bu_{i1}||^2\vartheta_i^2}} (1-||\bu_{i1}||^2)\vartheta_i\right)^2-\vartheta_i^2 \right]>0,\ \vartheta_i=\sqrt{\mathbb{E}\left(\hat{\mathbf{u}}_i^{\top} \mathbf{u}_i\right)^2}.
\end{align*}
\end{theorem}

Theorem \ref{th-nonsparse} reveals that our identification-recovery framework achieves superior performance compared to the direct use of local sample eigenvectors. The performance gain is quantified by the strictly positive quantity $\mathscr{C}(\mathcal A)$, which increases with the squared norms $\|\bu_i(\mathcal A)\|^2$ for $i\in [K]$.

We conclude this section by noting that all theoretical results discussed here can be naturally extended to the distributed PCA framework using correlation matrices.

\section{Simulation.}\label{sec:5}
In this section, we conduct simulation studies to evaluate the consistency of our estimator in comparison with some existing methods. To assess the performance of the proposed algorithm under different data distributions, we generate samples according to the model $\left\{\mathbf{x}_i=\boldsymbol{\Sigma}^{1 / 2} \mathbf{z}_i, i=1, \ldots, N\right\}$.
We consider two types of distributions for  $\mathbf{z}_i=\left\{z_{i j}\right\}: z_{i j} \sim \mathcal{N}(0,1)$ 
and $z_{i j} \sim \operatorname{Exp}(1)-1$.
For the population covariance matrix 
$\bSigma$, we examine two scenarios, each featuring two spiked eigenvalues.
\begin{itemize}
	\item [(1)] \textbf{Sparse case:}  $\bSigma_1=\diag\left( \frac{11}{16} \bf 1_4 \bf 1_4^{\top}  + \frac{1}{4} \bI_4 ,\frac{5}{8}  \bf 1_6 \bf 1_6^{\top}  + \frac{1}{4} \bI_6 , \mathbf I_{p/2-10}, \frac{1}{2} \mathbf I_{p/2}\right)$ and \(\bR_1 = \operatorname{diag}(\bSigma_1)^{-1/2} \bSigma_1 \operatorname{diag}(\bSigma_1)^{-1/2}\). The first two eigenvalues and eigenvectors are $\lambda_1(\bSigma_1)=4, \lambda_2(\bSigma_1) =3,\mathbf{u}_1^{\top}=\left[\mathbf{0}_4^{\top}, \tfrac{1}{\sqrt{6}} \mathbf{1}_6^{\top}, \mathbf{0}_{p-10}^{\top}\right], \mathbf{u}_2^{\top}=\left[\tfrac{1}{2} \mathbf{1}_4^{\top}, \mathbf{0}_{p-4}^{\top}\right]$ (for \(\bSigma_1\)) and
    $   \lambda_1(\bR_1) =4.57, \lambda_2(\bR_1) =3.2,
        \bv_1 = \bu_1,  \bv_2 = \bu_2 $ (for \(\bR_1\)).

	\item [(2)] \textbf{Mixed case:}  $\bSigma_2 = \diag( \bM, 0.7 \bf 1_4 \bf 1_4^{\top} + 0.2\bI_4, \mathbf I_{p/2-4})$, where
     \( \bM \in \mathbb{R}^{p/2 \times p/2} \)  is given by
\[
\bM = 
\begin{bmatrix}
a_2 \mathbf 1_{6} \mathbf 1_{6}^{\top} + (a_1-a_2) \bI_6  & a_3 \mathbf 1_{6} \mathbf 1_{p/2-6}^{\top} \\
a_3 \mathbf 1_{p/2-6} \mathbf 1_{6}^{\top}  & a_4 \mathbf 1_{p/2-6} \mathbf 1_{p/2-6}^{\top} + (a_1-a_4) \bI_{p/2-6},
\end{bmatrix}
\]
with $a_1 = 0.9, a_2 = 0.74,  
a_3 = (5- a_1 - 5a_2) \tfrac{\sqrt{3}}{\sqrt{p-12}}, 
a_4 = \tfrac{5-3a_3\sqrt{3p-12}-a_1}{p/2-7}.$
The first two eigenvalues and eigenvectors  are $\lambda_1(\bSigma_2)=5,  \lambda_2(\bSigma_2) =3,  \mathbf{u}_1^{\top}=\left[\sqrt{\tfrac{0.9}{6}}\mathbf{1}_6^{\top}, \sqrt{\tfrac{0.1}{p/2 - 6}}\mathbf{1}_{p/2 - 6}^{\top}, \mathbf{0}_{p/2}^{\top}\right],$ $\mathbf{u}_2^{\top}=\left[\mathbf{0}_{p/2-4}^{\top},\tfrac{1}{2} \mathbf{1}_4^{\top}, \mathbf{0}_{p/2-4}^{\top}\right]$ (for \(\bSigma_2\)) and $\lambda_1(\bR_2) =5.56, \lambda_2(\bR_2) =3.33, 
        \bv_1 =  \bu_1, \bv_2 = \bu_2$ (for \(\bR_2\)).

 \end{itemize}

\paragraph{Fixed local sample size.}
In this setting, we fix 
$p=3000$, and each machine is assigned a fixed local sample size. Specifically, one third of the machines have a sample size of 2400, another third have 2700, and the remaining third have 3000.
The tables below report the estimation error
$\rho=\left\|\check{\mathbf{U}}_K \check{\mathbf{U}}_K^{\top}-\mathbf{U}_K \mathbf{U}_K^{\top}\right\|_F$
, as the number of machines 
$m$ varies, using the sample covariance matrix (“Cov-PCA”) and the sample correlation matrix (“Cor-PCA”), respectively.
Each error value is averaged over 100 independent Monte Carlo repetitions, and the tuning parameter in our algorithm is set to $t=0.1.$
Compared with the methods of \citet{fan} and \citet{li2025two}, our approach consistently achieves lower error across a wide range of settings, with particularly substantial improvements when the number of machines $m$ is small or when the population eigenvectors are sparse.

\begin{table}[H]
	\centering
	\renewcommand{\arraystretch}{0.6}
	\begin{tabular}{l|ccc|ccc}
		\hline 
		& \multicolumn{3}{|c}{Cov-PCA} & \multicolumn{3}{|c}{Cor-PCA}  \\ \hline
		$z_{ij} \sim \mathcal{N}(0,1)$ &  $\rho$ & $\rho^{Fan}$ & $\rho^{Li}$ & $\rho$ & $\rho^{Fan}$ & $\rho^{Li}$ \\ \hline
		$m=3$     & 0.0263 & 0.7591 & 0.7354 & 0.0370 & 0.8578 & 0.8232 \\ 
		$m=30$     &  0.0069 & 0.2583 & 0.2435 & 0.0062 & 0.3001 & 0.2760 \\ 
		$m=90$    & 0.0054 & 0.1500 & 0.1412 & 0.0035 & 0.1747 & 0.1601 \\  
		$m=300$    &  0.0021 & 0.0824 & 0.0775 & 0.0019 & 0.0962 & 0.0879 \\
		\hline
		$z_{ij} \sim \mathrm{Exp}(1) - 1$ &  $\rho$ & $\rho^{Fan}$ & $\rho^{Li}$ & $\rho$ & $\rho^{Fan}$ & $\rho^{Li}$ \\ \hline
		$m=3$    & 0.0317 & 0.7609 & 0.7365 & 0.0292 & 0.8581 & 0.8238 \\  
		$m=30$   &  0.0098 & 0.2590 & 0.2438 & 0.0062 & 0.3001 & 0.2760 \\ 
		$m=90$   &  0.0048 & 0.1505 & 0.1412 & 0.0036 & 0.1748 & 0.1601 \\ 
		$m=300$   &   0.0025 & 0.0828 & 0.0776 & 0.0019 & 0.0962 & 0.0880 \\  \hline
	\end{tabular}
	\caption{ Estimation error under the sparse case.}
	\label{table:f1}
\end{table}

\begin{table}[H]
	\centering
	\renewcommand{\arraystretch}{0.6}
	\begin{tabular}{l|ccc|ccc}
		\hline 
		& \multicolumn{3}{|c}{Cov-PCA} & \multicolumn{3}{|c}{Cor-PCA}  \\ \hline
		$z_{ij} \sim \mathcal{N}(0,1)$ &  $\rho$ & $\rho^{Fan}$ & $\rho^{Li}$ & $\rho$ & $\rho^{Fan}$ & $\rho^{Li}$ \\ \hline
		$m=3$   & 0.4343 & 1.0274 & 0.9740 & 0.4139 & 0.9707 & 0.9263 \\ 
		$m=30$   & 0.1906 & 0.3878 & 0.3357 & 0.1784 & 0.3574 & 0.3178 \\ 
		$m=90$   & 0.1100 & 0.2282 & 0.1954 & 0.1046 & 0.2095 & 0.1848 \\  
		$m=300$ & 0.0612 & 0.1258 & 0.1074 & 0.0582 & 0.1154 & 0.1016 \\     \hline
		$z_{ij} \sim \mathrm{Exp}(1) - 1$ &  $\rho$ & $\rho^{Fan}$ & $\rho^{Li}$ & $\rho$ & $\rho^{Fan}$ & $\rho^{Li}$ \\ \hline
		$m=3$   &  0.4246 & 1.0315 & 0.9769 & 0.4040 & 0.9689 & 0.9250 \\  
		$m=30$   & 0.1900 & 0.3917 & 0.3376 & 0.1753 & 0.3583 & 0.3186 \\ 
		$m=90$    &  0.1129 & 0.2298 & 0.1958 & 0.1041 & 0.2096 & 0.1848 \\  
		$m=300$  & 0.0614 & 0.1265 & 0.1073 & 0.0581 & 0.1153 & 0.1014 \\ \hline
	\end{tabular}
	\caption{Estimation error under the mixed case.}
	\label{table:f2}
\end{table}

\begin{figure}[H]
    \begin{subfigure}{0.48\textwidth}
        \centering
\includegraphics[width=\textwidth]{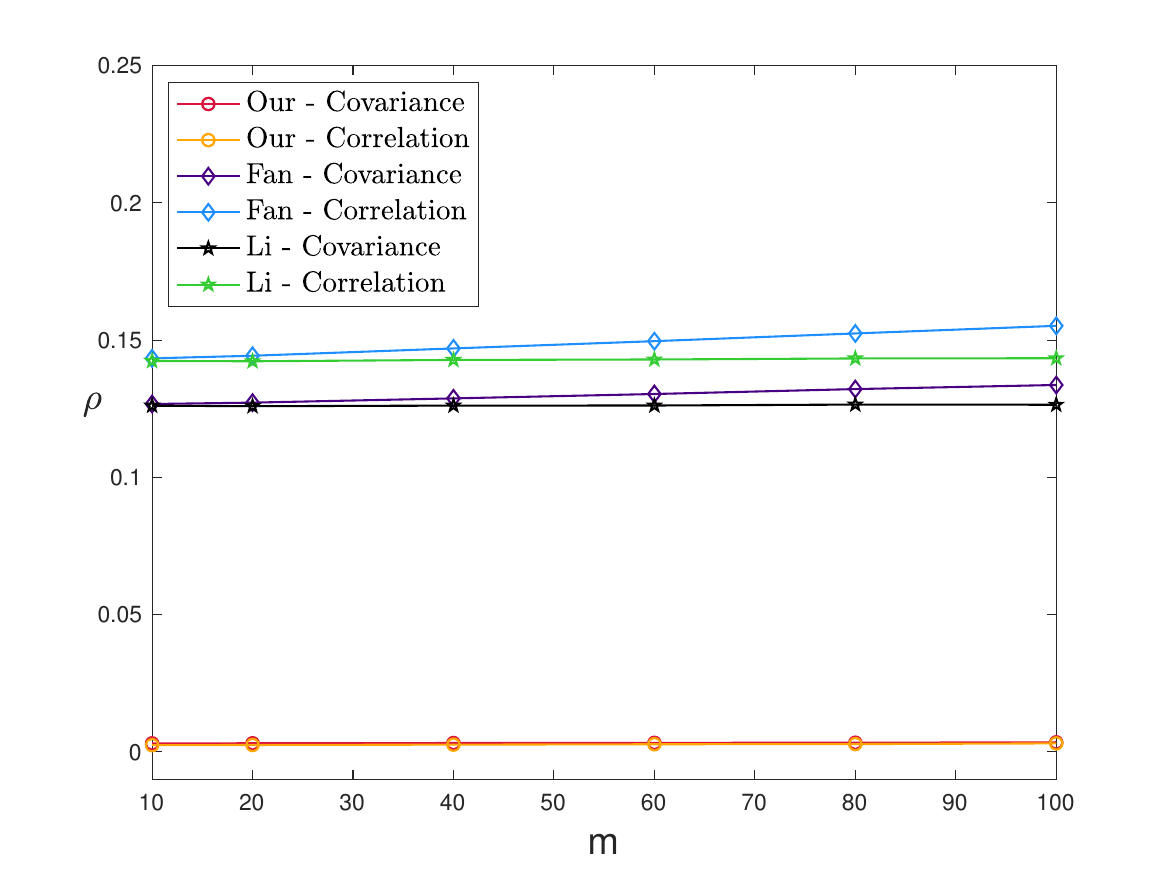} 
        \caption{$z_{i j} \sim \mathcal{N}(0,1).$}
\end{subfigure}
           \begin{subfigure}{0.48\textwidth}
        \centering
   \includegraphics[width=\textwidth]{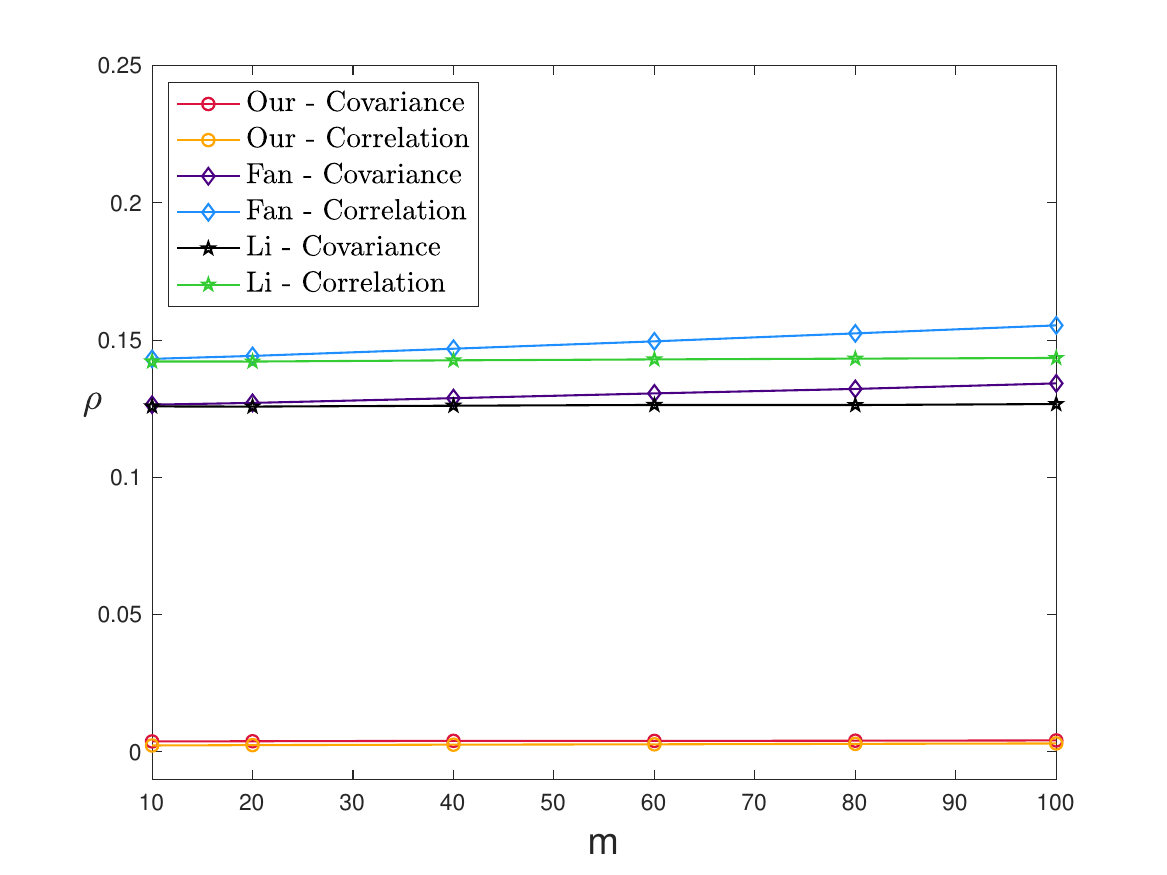} 
        \caption{$z_{i j} \sim \operatorname{Exp}(1)-1.$}
    \end{subfigure}
    \caption{Estimation error under the sparse case.}
    \label{fig:fixed Ns}
\end{figure}
\begin{figure}[H]
    \begin{subfigure}{0.48\textwidth}
        \centering
\includegraphics[width=\textwidth]{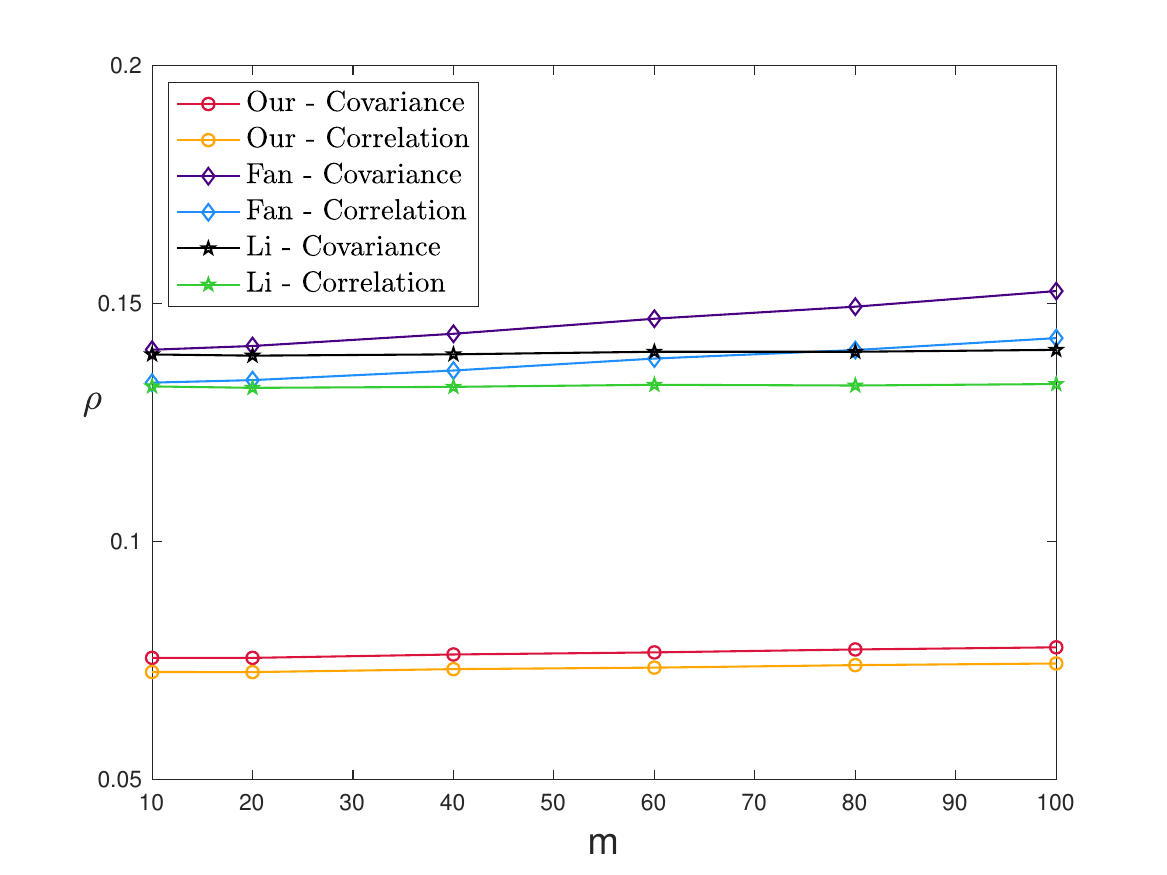} 
        \caption{$z_{i j} \sim \mathcal{N}(0,1).$}
\end{subfigure}
           \begin{subfigure}{0.48\textwidth}
        \centering
   \includegraphics[width=\textwidth]{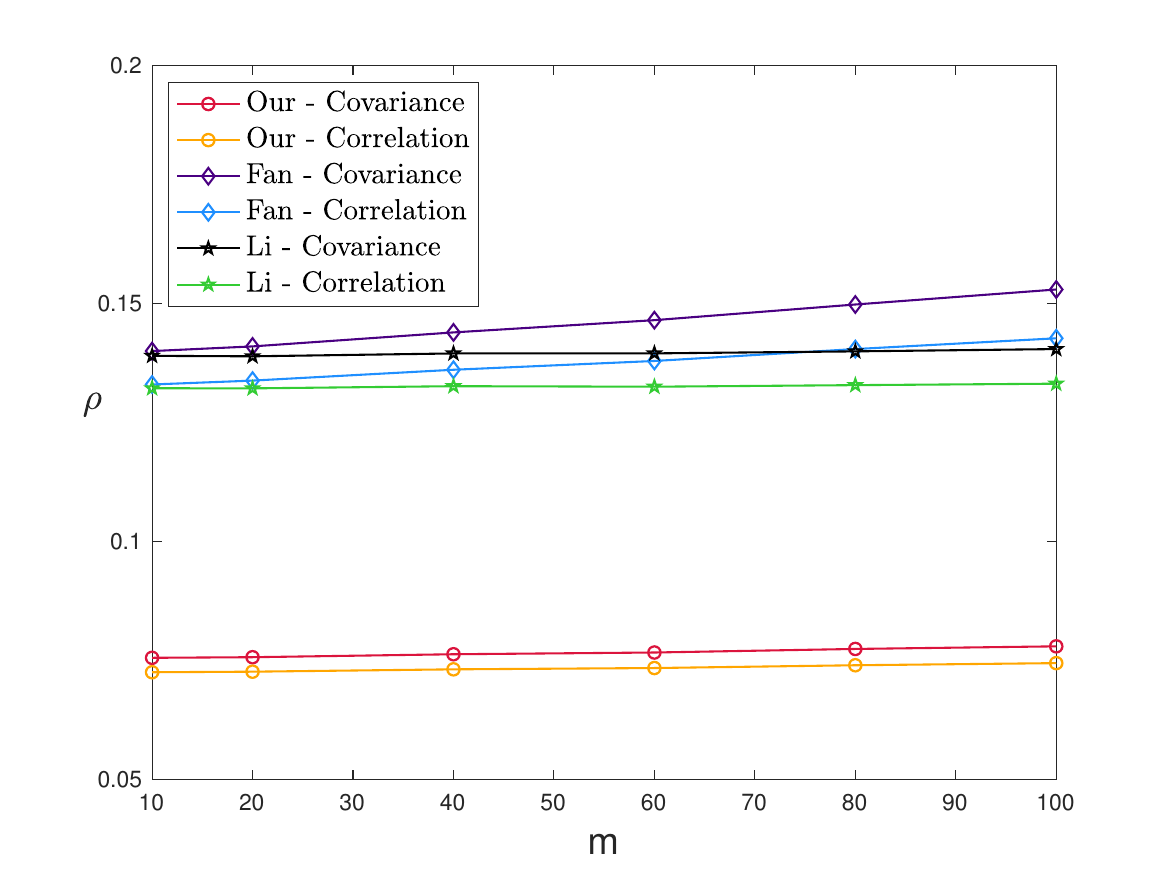} 
        \caption{$z_{i j} \sim \operatorname{Exp}(1)-1.$}
    \end{subfigure}
    \caption{Estimation error under the mixed case.}
    \label{fig:fixed Nm}
\end{figure}

\paragraph{Fixed total sample size.} 
In this setting, we set $p=3000$ and fix the total sample size at $N=3 \times 10^5$.
The data are evenly distributed across $m$ machines, and we investigate how the estimation error varies with different values of $m$. 
As shown in Figures \ref{fig:fixed Ns}-\ref{fig:fixed Nm}, our method consistently outperforms existing approaches under all considered scenarios—whether using the sample covariance matrix (Cov-PCA) or the sample correlation matrix (Cor-PCA)—across different levels of sparsity (sparse or mixed) and varying numbers of machines $m$.

\section{Real data analysis.}\label{sec:6}
In this section, we apply the distributed PCA algorithms to the  MiniBooNE dataset 
which is collected from tabular data benchmark  (\url{https://huggingface.co/datasets/inria-soda/tabular-benchmark}). The MiniBooNE dataset \citep{roe2005boosted}  originates from the neutrino experiments conducted at the Fermi National Accelerator Laboratory and is used to distinguish electron neutrinos (signal) from muon neutrinos (background). It consists of $72{,}998$ samples, each with $50$ numerical features. Detailed information about the dataset is available at: \url{https://huggingface.co/datasets/inria-soda/tabular-benchmark/blob/main/clf_num/MiniBooNE.csv}.

\begin{figure}[H]
    \centering  \includegraphics[width=\linewidth]{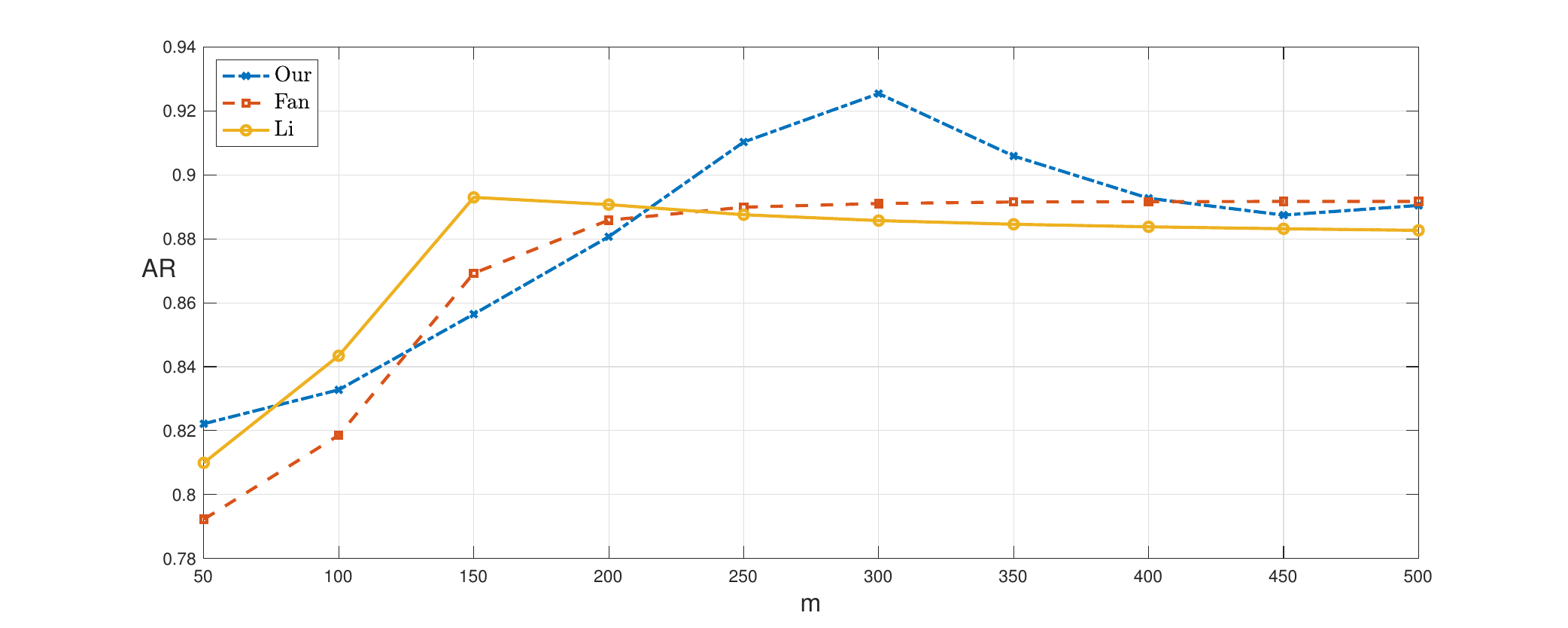}
    \caption{Average AR  based on 1000 replications for different number of machines $m$.}
     \label{fig:mini}
\end{figure}

To evaluate the performance of eigenspace estimation, we split the dataset into 70\% for training and 30\% for testing. Using the training data, we estimate the top-$K$ eigenmatrix $\mathbf{U}_K$ with $K=3$, and then apply the resulting projection to the test data to compute the Average Information Preservation Ratio (AR), defined as
$
\operatorname{AR}(\bU_K) = \sum_{i } \| \bU_K \bU_K^{\top} \by_i \|_2^2 / \sum_{i } \| \by_i \|_2^2.$
This metric quantifies how well the estimated subspace preserves the informative structure of the data, with higher AR values indicating more effective estimation. For estimating $\mathbf{U}_K$, we apply different distributed PCA algorithms while varying the number of machines $m$. In our method, we implement Algorithm 1 
based on sample covariance matrix and experiment with the tuning parameter $t =0.005$, and compare with \citet{fan} and \citet{li2025two}. Figure~\ref{fig:mini} presents the average AR values over 1000 replications across different number of machines $m$. As shown, our method generally achieves higher AR scores, demonstrating superior performance in preserving information under distributed settings.

\bigskip
\begin{center}
{\large\bf SUPPLEMENTARY MATERIAL}
\end{center}

\textbf{Supplementary Material of ``Debiased distributed PCA under high dimensional spiked model "}: This supplementary material contains the proofs of Theorems \ref{th-main}-\ref{th-nonsparse} and an alternative proof for the consistency of $\tilde \bU_{K, Fan}$ under weaker conditions.

\bibliographystyle{apalike}
\bibliography{sample}
\end{document}